\documentclass[useAMS,usenatbib]{mnras}
\usepackage{graphicx}
\usepackage{amsbsy,amsmath}
\usepackage{graphics,graphicx}
\usepackage{subfloat}
\usepackage[compatibility=false]{caption}
\usepackage[skip=0.1pt]{subcaption}
\usepackage{ae,aecompl}
\usepackage[compact]{titlesec}
\usepackage{notoccite}
\usepackage{caption}
\usepackage{subcaption}
\usepackage{amsbsy,amsmath}
\usepackage{graphics,graphicx}
\usepackage{longtable}
\usepackage[latin1]{inputenc}
\usepackage{amsmath}
\usepackage{amsfonts}
\usepackage{amssymb}
\usepackage{makeidx}
\usepackage{float}
\restylefloat{figure}
\usepackage{stfloats}
\usepackage{epsfig}
\usepackage{graphicx}
\usepackage{caption}
\usepackage{color}
\usepackage{tabularx}
\usepackage{natbib}
\usepackage{multicol}
\usepackage{gensymb}
\usepackage[toc]{appendix}

\title{A rare case of FR I interaction with a hot X-ray bridge in the A2384 galaxy cluster}

\author[Parekh et al.]{V. Parekh$^{1,2}$\thanks{E-mail: viralp@ska.ac.za}, 
T. F. Lagan\'{a}$^{3}$, 
K. Thorat$^{1,2}$, 
K. van der Heyden$^{4}$, 
A. Iqbal$^{5}$ \newauthor and 
F. Durret$^{6}$
\\
$^{1}$Department of Physics and Electronics, Rhodes University, Grahamstown, Republic of South Africa\\
$^{2}$South African Radio Astronomy Observatory (SARAO), Cape Town, Republic of South Africa \\
$^{3}$N\'ucleo de Astrof\'\i sica, Universidade Cruzeiro do Sul/Universidade Cidade de S\~ao Paulo, \\
    R. Galv\~ao Bueno 868, Liberdade, S\~ao Paulo, SP, 01506-000, Brazil\\
$^{4}$ Department of Astronomy, University of Cape Town, Cape Town, Republic of South Africa \\
$^{5}$ Raman Research Institute, Sadashiv Nagar, Bangalore, India\\
$^{6}$Sorbonne Universit\'e, CNRS, UMR~7095, Institut d'Astrophysique
de Paris, 98bis Bd Arago, 75014, Paris, France
}

\date{MNRAS accepted.}

\pubyear{2019}

\begin{document}

\label{firstpage}
\pagerange{\pageref{firstpage}--\pageref{lastpage}}
\maketitle

\begin{abstract}

Clusters of varying mass ratios can merge and the process significantly disturbs the cluster environments and alters their global properties. Active radio galaxies are another phenomenon that can also affect cluster environments. Radio jets can interact with the intra-cluster medium (ICM) and locally affect its properties. Abell~2384 (hereafter A2384) is a unique system that has a dense, hot X-ray filament or bridge connecting the two unequal mass clusters A2384(N) and A2384(S). The analysis of its morphology suggests that A2384 is a post-merger system where A2384(S) has already interacted with the A2384(N), and as a result hot gas has been stripped over a $\sim 1$ Mpc region between the two bodies. We have obtained its 325 MHz GMRT data, and we detected a peculiar FR I type radio galaxy which is a part of the A2384(S). One of its radio lobes interacts with the hot X-ray bridge and pushes the hot gas in the opposite direction. This results in displacement in the bridge close to A2384(S). Based on {\it Chandra} and XMM-{\it Newton} X-ray observations, we notice a temperature and entropy enhancement at the radio lobe-X-ray plasma interaction site, which further suggests that the radio lobe is changing thermal plasma properties. We have also studied the radio properties of the FR I radio galaxy, and found that the size and radio luminosity of the interacting north lobe of the FR I galaxy are lower than those of the accompanying south lobe.   
\end{abstract}

\section{Introduction}
\par The formation and evolution of galaxy clusters most likely results from complex mergers of multi-component systems. Cluster mergers are highly dynamical processes \citep{2002ASSL..272....1S}, and numerous, heterogeneous merger episodes \citep{2006A&A...446..417F,2006MNRAS.373..881P} are considered to be the most natural physical processes for disturbing both their cool-core and relaxed structure. Even in the present age, clusters are constantly growing by matter accretion at the nodes of large scale filaments of galaxies \citep{2005Natur.435..629S,2008SSRv..134..311D,2014MNRAS.438.3465T,2017ApJ...844...25B}. 
\par Recently, X-ray observations of clusters have shown a variety of features that are due to either minor or major mergers, such as cold fronts, gas sloshing, substructures, shock edges, etc.,  \citep{2013ApJ...779..112N,2013A&A...549A..19W,2015A&A...575A.127P}. One of the most evident is a bimodal pair of galaxy clusters, with a hot, dense Mpc scale extended gaseous region between the two clusters taking the form of a filament or bridge \citep{2008A&A...482L..29W,2016A&A...593L...7A,2017MNRAS.470.3742P}. The imprints of the merger process depend on the masses of the subsystems and on the impact parameter. 
These bimodal clusters are ideal probes to study the different evolutionary epochs in the formation of clusters, to understand how the merger shocks propagate and dissipate kinetic energy during the cluster merging process, and to explore the intra-cluster medium (ICM) \citep{2003ApJ...593..599R,2010PASJ...62..335A,2013A&A...550A.134P}. \cite{2010A&A...509A..86M} have detected diffuse radio halos among the close pair of each of the clusters A399-A401, which is considered to be a sporadic detection. This radio detection is important to study the origin of diffuse radio halos in connection with the evolutionary stage of the merger. Optical observations also support the scenario of a galaxy filament tracing the hot gaseous bridge between two merging clusters \citep{2011A&A...525A..79M}.
\par Joint radio and X-ray analyses show many interesting phenomena taking place in the interaction between radio galaxies and the ICM \citep{2002ASSL..272..163F,2003AJ....125.1635B,2011AJ....141...88W,2017ApJ...844...78P}. 
For example, {X-ray observations have shown cavities in the ICM} of a number of relaxed and cool core clusters. The radio observations of these clusters reveal that these cavities are the result of the interaction of tailed radio galaxies with the X-ray emitting ICM, where the radio bubbles or lobes fill these X-ray cavities \cite[and references therein]{2007ARA&A..45..117M,2012NJPh...14e5023M}. The powerful AGN launch radio jets which create cavities and jet outbursts, that are considered to be a feedback mechanism regulating the cooling flow in cool core clusters \citep{1994ARA&A..32..277F,2006PhR...427....1P,2012ARA&A..50..455F}. The two types of plasmas, both thermal and non-thermal, do not mix very well and the jets inflate bubbles in the ICM. Another widely seen example is the ICM exerting pressure on fast moving radio galaxies inside a dense cluster environment. As a result, a wide range of distorted and asymmetric extended morphologies is visible among radio galaxies. In this case, the radio structure in which lobes or plumes are not aligned with the central galaxy can be due to the bending of jets, generally known as bent-tail (BT) radio galaxies \citep{2014AJ....148...75D}. Bent tail source host galaxies are usually non-central to the cluster.

\par {AGN-powered radio galaxies can be classified on a morphological basis into Fanaroff-Riley Class I and II (FR I/FR II henceforth) \citep{1974MNRAS.167P..31F}. The classification is defined by the ratio of the distance between the brightest points in the source to the total extent of the source. If this ratio is greater than 0.5 for a given source, the source would be classified as an FR II source; otherwise it would be put in the FR I class. Sources belonging to the FR I class have morphology typified by a comparatively bright core and diffuse, plume-like jets/lobes. FR II sources, on the other hand, show the brightest emission at the ends of their extent (these are called `hotspots').  Bent-tail radio sources are predominantly FR I sources that are rapidly moving through the dense ICM and so become distorted}.   

\par In this paper, we studied the uncommon phenomenon of the interaction between the radio lobe and the extended thermal X-ray emission in the bimodal A2384 galaxy cluster, where the radio lobe displaces the hot filamentary gas on a large scale. We also studied the thermodynamical properties of the X-ray bridge to understanding the impact of radio emission on the hot thermal gas. Based on the Simbad\footnote{http://simbad.u-strasbg.fr/simbad/} and NED databases\footnote{https://ned.ipac.caltech.edu/}, the redshift of A2384 is $z$ = 0.0943 and \cite{2011A&A...525A..79M} classified it as {Abell richness class 1} and BM-type II-III. We selected the A2384 galaxy cluster from \cite{2013A&A...550A.134P} and it was detected in the Planck observation and identified as PLCKESZ G033.46-48.43 \citep{2014A&A...571A..29P}. This cluster shows a very complex X-ray morphology and a dense ICM between the pair of clusters. The selection of this cluster includes the availability of high-quality {\it Chandra} and XMM-{\it Newton} X-ray data and of optical spectroscopic data to identify its member galaxies. A2384 was previously studied in X-rays by \cite{1982ApJ...258..434U} with the Einstein satellite, by \cite{1996rftu.proc..587H} with ROSAT HRI and by \cite{2011A&A...525A..79M} with XMM-{\it Newton}. In all these X-ray observations, the two clusters were found to be separated by a projected $\sim$ 1 Mpc X-ray bridge. {Further}, \cite{2004ApJ...613...95C} and \cite{2014A&A...570A..40P} performed weak lensing and detailed optical analysis of A2384.   

\par This paper is organised as follows; Section~2 gives a note on the serendipitous discovery of FR I radio galaxy, Section~3 describes the X-ray and radio observations and their data reduction procedures. Section~4 presents our images and results. Section~5 shows the X-ray spectral analysis. Section~6 is the discussion and Section~7 gives the conclusions. We assumed $H_{0}$ = 70 km s$^{-1}$ Mpc$^{-1}$ $\Omega_{M}$ = 0.3 and $\Omega_{\Lambda}$ = 0.7 throughout the paper. At redshift $z$ = 0.0943, 1$''$ corresponds to 1.75 kpc.

\section{Serendipitous discovery of A2384 FR I radio galaxy}
\par {Diffuse radio sources in the form of radio halos and relics of the size of $\sim$ 1 Mpc are mostly detected in merging and massive clusters \citep{2019SSRv..215...16V,2012A&ARv..20...54F}. We studied A2384 for two main reasons: (1) to detect diffuse radio emission in the form of a radio halo associated with A2384(N), as it shows disturbed X-ray morphology and (2) to achieve a sensitive low frequency radio observation of the A2384 X-ray bridge region, to detect the magnetic field and relativistic particles in the region between the two merging clusters, as recently detected in the merging clusters A399-A401 with  LOFAR data \citep{2019Sci...364..981G}. Early radio observations of A2384 were not sensitive enough for our studies. From our newer high quality GMRT data, we discovered a peculiar FR I radio galaxy associated with the A2384 system. In this work, we evaluated the radio properties of a FR I galaxy and its interaction with the ICM of the A2384 bridge.    }   

\section{X-ray and Radio observations}

\subsection{Chandra data analysis}
\par We retrieved {\it Chandra} archival data  (obsid 4202, date of observation 2002-11-18, and total exposure time $\sim$ 31,445 sec) to study A2384. We used high-resolution {\it Chandra} data for both imaging and spectral analysis purposes. For the data reduction, we followed the standard X-ray data analysis procedure. We processed the {\it Chandra} data with the CIAO (v 4.11 and CALDB 4.8.2) software. In the data reduction steps, we first used the \verb"chandra_repro" task to reprocess ACIS imaging data, followed by removing high background flares (3$\sigma$ clipping) with the task \verb"lc_sigma_clip". All filtered event files ({exposure time $\sim$ 31,254 sec} after flare removal) included the 0.3--7 keV broad energy band and 1.968$''$ pixel binning. We detected and removed point sources around the cluster and bridge regions using the \verb"wavdetect" and \verb"dmfilth" tasks, respectively. Finally, we divided the count image by the exposure map and generated the flux image (photon~cm$^{-2}$~s$^{-1}$). We also subtracted the background from the flux image using Chandra blank sky files\footnote{http://cxc.harvard.edu/ciao/threads/acisbackground/}. 

\subsection{XMM-Newton Data Reduction}
\label{sec:XMM}

\par A2384 was observed on 2004 October 28th for $\sim$ 26~ks for MOS and $\sim$ 19.4~ks for pn ({ XMM-{\it Newton} observation id. 0201902701}). Prime full frame mode was used for the three cameras (extended mode for pn) with thin filters. 
The data reduction was done with SAS version 16.1.0 (July 2017)  and calibration files updated to 2019 April.  In order to filter background flares, we applied a 2$\sigma$-clipping procedure using the { light curves in the [10--14]\,keV energy band}. The resulting ``cleaned'' exposure times are 15.62 ks for MOS cameras and  8.17 ks for pn, respectively. To take into account the background contribution for each detector, we obtained a background spectrum in an outer annulus of the observation in the 10-12 keV energy band. We compared these spectra with the blank sky obtained by \citet{ReadPonman03} in the same region and energy band. Then, we rescaled the observation background to the blank sky background to obtain a normalisation parameter that will be used in the spectral fits. Point sources were detected by visual inspection, confirmed in the High Energy Catalogue 2XMMi Source, and excluded from our analysis. 
\par For XMM-{\it Newton} imaging purposes, we used the ``IMAGES'' script\footnote{https://www.cosmos.esa.int/web/xmm-newton/images} to generate combined MOS1, MOS2 and pn smoothed images. This script works as a pipeline to automatically filter out high background flares, bad pixels and columns, apply exposure corrections and finally combine MOS and pn data {for the specified energy bands}. We then smoothed these final images.

\subsection{Radio data analysis}
\par We observed the A2384 cluster with the upgraded GMRT band 3 (250-500 MHz) in January 2019 for a total duration of $\sim$ 6 hours (P.I. V. Parekh, obsid. 28$\_$087). {This u-GMRT observation delivered both the narrow band (325 MHz central frequency with 32 MHz bandwidth) and wideband (400 MHz central frequency with 200 MHz bandwidth) data}.  The narrow band data were processed with a fully automated pipeline based on the SPAM package \citep{2009A&A...501.1185I,2014arXiv1402.4889I}, which includes direction-dependent calibration, modelling and imaging to suppress mainly ionospheric phase errors. Briefly, the pipeline consists of two parts: (1) a \emph{pre-processing} part that converts the raw data from individual observing sessions into pre-calibrated visibility data sets for all observed pointings, and (2) a \emph{main pipeline} part that converts pre-calibrated visibility data per pointing into Stokes I continuum images. The flux scale is set by calibration on 3C286 using the models from \citet{2012MNRAS.423L..30S}. We used different robust weighting and outer tapering values in the IMAGR task to make high and low resolution images. These images are primary beam corrected. More details on the processing pipeline and characteristics of the data products are given in the paper on the first TGSS alternative data release \citep[ADR1]{2017A&A...598A..78I}. Our wide band data reduction process is ongoing and we aim to report the corresponding results in our next paper.

\section{Images}

\par
\begin{figure*}
    \centering
    \begin{subfigure}[t]{0.50\textwidth}
        \includegraphics[width=1\textwidth]{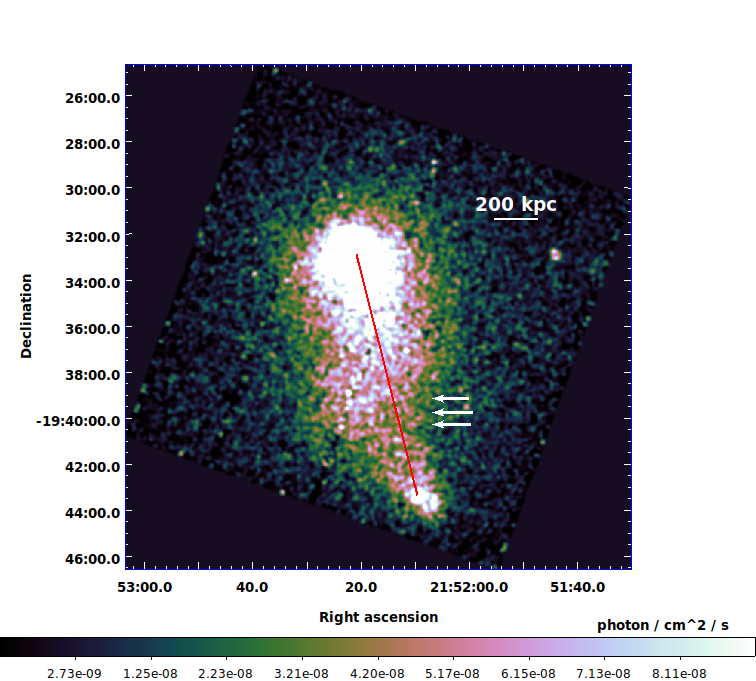}
        \caption{}
        \label{rfidtest_xaxis1}
    \end{subfigure}
    \begin{subfigure}[t]{0.45\textwidth}
        \includegraphics[width=1\textwidth]{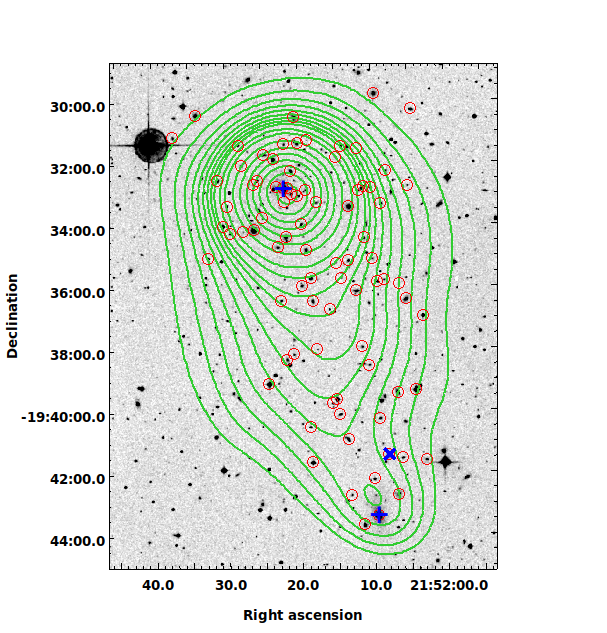}
        \caption{}
        \label{rfidtest_yaxis2}
        \end{subfigure}
    \caption{(a) A2384 exposure corrected and background subtracted {\it Chandra} X-ray image. The white arrows show the notch or displacement location in the X-ray bridge. The red line shows the projected distance between two X-ray peaks. (b) A2384 DSS optical image in grey color. The red circles show the member galaxies from \citet{2011A&A...525A..79M}. The green contours show the {\it Chandra} smoothed ($\sigma$ = 15$''$) X-ray emission. BCGs of A2384(N) and A2384(S) are marked with blue `+' symbols. The LEDA 851827 galaxy is marked with an blue `x'.}
    \label{Xray_img}        
\end{figure*}

 A2384 is very peculiar cluster system \citep{2011A&A...525A..79M,2014A&A...570A..40P} that displays a dense X-ray filament ($\sim$ 700 kpc size) between A2384(N) (North cluster) and A2384(S) (South cluster) (Fig.~\ref{Xray_img}(a)). A2384 is similar to the A3017 \citep{2017MNRAS.470.3742P} and A222-A223 \citep{2008A&A...482L..29W} cluster systems, which also display  X-ray filament between two sub-clusters. The optical data analysis of the galaxy distribution \citep{2011A&A...525A..79M}, at all magnitude limits, also shows a very elongated structure along the north-south axis (Fig.~\ref{Xray_img}(b)). A total of 56 cluster members have been identified, and both the X-ray gas and galaxy distribution approximately define the same bimodal structure. Recently, \cite{2014A&A...570A..40P} have used 2dF and EFOSC2 spectroscopic data of A2384 and detected another substructure in the east of the central cluster region, which is falling onto the cluster and makes this system even more puzzling. The properties of A2384 are given in Table \ref{sample}. The X-ray observations show that A2384 is elongated from north to south with a physical size of $\sim$ 1.5 Mpc and an X-ray peak to peak distance of $\sim$ 1.1 Mpc. In the {\it Chandra} X-ray map, there is a displacement visible in the X-ray bridge at the distance of $\sim$ 800~kpc from the X-ray peak of A2384(N). At this position, the X-ray gas becomes distorted and bent towards the east. We have marked this by white arrows in Fig.~\ref{Xray_img}(a), and the red line in the same figure shows the projected distance between A2384(N) and A2384(S). The distribution of gas in the bridge is broader (7$'$) up to this displacement and then becomes narrower (4$'$), and connects to A2384(S). Such a displacement is uncommon and is probably noticed here for the first time on such a large scale. 

\par
\begin{center}
\begin{figure*}
    \centering
    \begin{subfigure}[t]{0.50\textwidth}
        \includegraphics[width=1\textwidth]{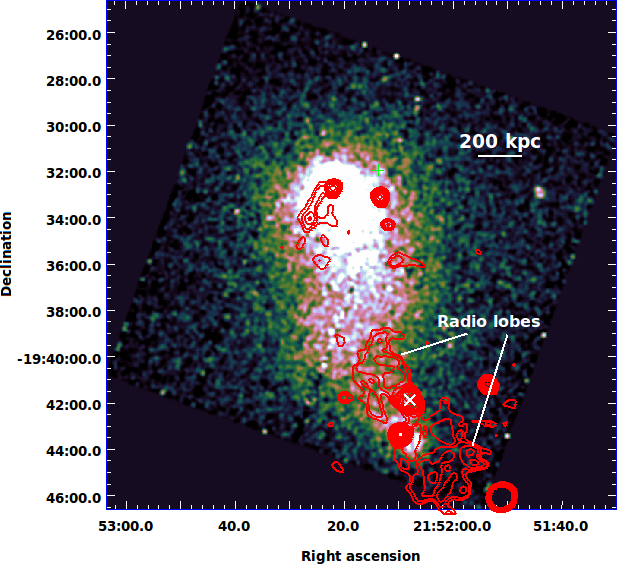}
        \caption{}
        \label{rfidtest_xaxis1}
    \end{subfigure}
    \begin{subfigure}[t]{0.40\textwidth}
        \includegraphics[width=1\textwidth]{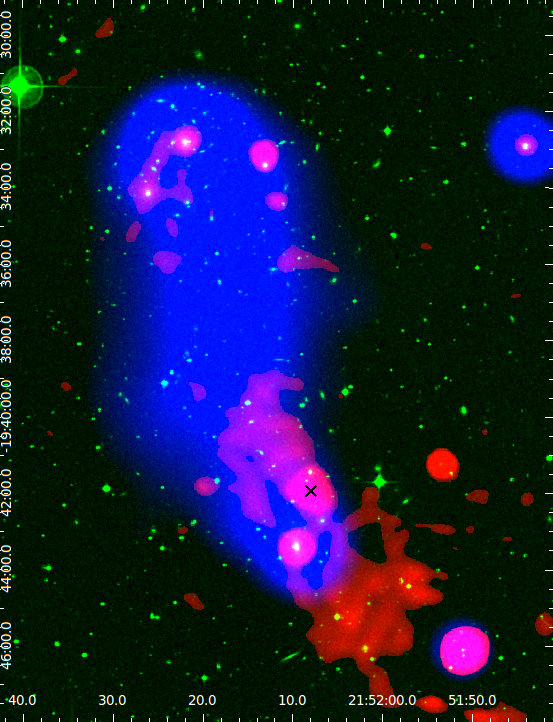}
        \caption{}
        \label{rfidtest_yaxis2}
        \end{subfigure}
    \caption{(a) A2384 {\it Chandra} X-ray color image. The red contours show the GMRT 325 MHz observation. The contours increase in steps of $\sqrt{2}$. The first contour is drawn at 3$\sigma$.  (b) Red is the radio GMRT 325 MHz (3$\sigma$) image, green is the DSS optical image, and blue is the XMM-{\it Newton} X-ray image of A2384. In both (a) and (b) images the radio galaxy LEDA 851827 is marked with an `x'. In the GMRT radio images, the beam size is 25$''$ $\times$ 25$''$ and 1$\sigma$ = 0.8 mJy beam$^{-1}$.}
    \label{Xray_radio_img}        
\end{figure*}
\end{center}

 We plotted the 325 MHz GMRT radio contours on the {\it Chandra} X-ray image in Fig.~\ref{Xray_radio_img}(a). The FR I radio galaxy located 100$''$ north of the BCG (LEDA 67523) of A2384(S) is detected. We classified this galaxy as FR I is based on its morphology. {Two lobes} extended in the NE-SW direction are associated with the FR I radio galaxy. From the Simbad astronomical database, we found that the counterpart of the FR I radio galaxy is LEDA 851827. The optical position of LEDA 851827 (J2000 RA:21h52m08.1s; Dec:-19d41m27s) matches well its radio counterpart within the 25$''$ beam of the 325~MHz GMRT image. The spectroscopic redshift of this galaxy is $\sim$ 0.09447; hence, this galaxy is a member of the A2384 cluster. This further supports the observational evidence of the interaction between the NE radio lobe of LEDA 851827 and the X-ray bridge. { \cite{2011A&A...525A..79M} also identified this galaxy as a cluster member of A2384 in their optical catalogue}. {This host radio galaxy also has a counterpart at 1.4 GHz} (NVSS 215208-194147) \citep{1998AJ....115.1693C}. However, both radio lobes are absent in the NVSS observation. The physical size of this radio galaxy (from the top to the bottom of the radio lobes) is $\sim$ 830~kpc. The X-ray and radio images show that the NE lobe is displacing the X-ray gas in the east direction and is causing the discontinuity in the X-ray emissivity.
We estimated, by visual inspection, that the total gas displacement could be $\sim$ 400 kpc in the east direction from the projected distance between A2384(N) and A2384(S). We also detected a tailed radio galaxy which could be associated with the BCG (LEDA 190740) of A2384(N). Its tail is extending $\sim$ 1.35$'$ (141 kpc) towards the south-east direction.  
\par In the GMRT 325 MHz map, we have an rms of 0.8 mJy beam$^{-1}$ and a beam size of 25$''$ $\times$ 25$''$. We calculated (1) the north lobe size is $\sim$ 165$''$ $\times$ 134$''$ and its total integrated flux density $\sim$ 112 $\pm$ 12 mJy beam$^{-1}$, (2) the south lobe size is $\sim$ 240$''$ $\times$ 220$''$ and its total integrated flux density $\sim$ 173 $\pm$ 18 mJy beam$^{-1}$, and (3) for the entire FR~I radio galaxy, the flux density is $\sim$ 421 $\pm$ 43 mJy beam$^{-1}$. 

\par In the ACIS {\it Chandra} data, there are CCD gaps in the bridge region that
could interfere in the analysis. Thus, we also prefer to consider XMM-{\it Newton} data for A2384, which have different CCD geometries, better sensitivity and larger FOV (field of view) as compared to {\it Chandra}. We have shown MOS1, MOS2 and pn combined smoothed images in Fig.~\ref{Xray_radio_img}(b) together with radio and DSS optical images. We confirmed that the X-ray properties of the bridge, as mentioned above, in the XMM-{\it Newton} image are the same as in the {\it Chandra} observation and there are no observational artefacts which caused the displacement in the X-ray bridge.

\begin{table*}
\centering
\caption{Properties of A2384. For A2384(N) and (S), the positions and redshifts $z$ correspond to the two BCGs. $M_{200}$ taken from \protect \cite{2014A&A...570A..40P}, $M_{500}$ and $L_{500}$ taken from the MCXC catalogue and $M_{gas}$ from this work. $M_{SZ}$ is the Sunyaev--Zel'dovich mass proxy taken from the \protect \cite{2014A&A...571A..29P}.}
\begin{tabular}{ccccccccccccccc}
\hline
\hline
Cluster & RA(J2000) & DEC(J2000) & $z$ & $M_{200}$ & $M_{500}$ & $M_{gas}$& $M_{SZ}$ &$L_{500}$\\
 components   & h~~m~~s & d~~m~~s & & 10$^{15}$ $M_{\odot}$ & 10$^{14}$ $M_{\odot}$ &10$^{13}$ $M_{\odot}$ & 10$^{14}$ $M_{\odot}$&10$^{44}$ erg s$^{-1}$\\
\hline
A2384(N) &   21 52 21.9 &  -19 32 48.6 &0.092   &1.84   &1.30   &1.42  &-&-\\ 
A2384(S) &   21 52 09.5 &  -19 43 23.7 &0.096   &1.17   &0.78   &0.86  &-&-     \\
A2384(Total) & - &-                    &0.0943  &2.34   &2.61   &2.84  &4.07 &1.66\\
\hline
\end{tabular}
\label{sample}
\end{table*}

\subsection{Surface brightness profile}
\begin{center}
\begin{figure*}
    \centering
    \begin{subfigure}[t]{0.50\textwidth}
        \includegraphics[width=1\textwidth]{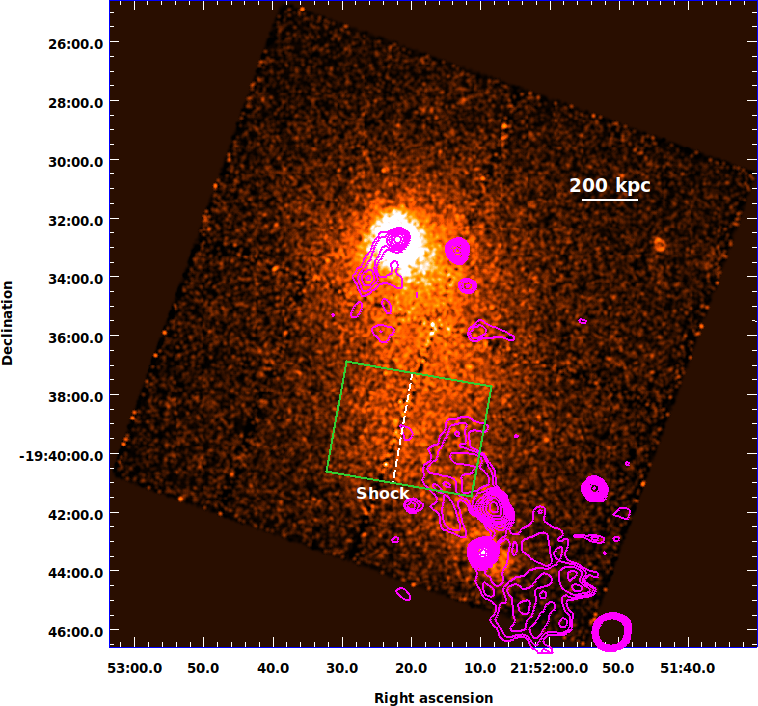}
        \caption{}
        \label{rfidtest_xaxis1}
    \end{subfigure}
    \begin{subfigure}[t]{0.49\textwidth}
        \includegraphics[width=1\textwidth]{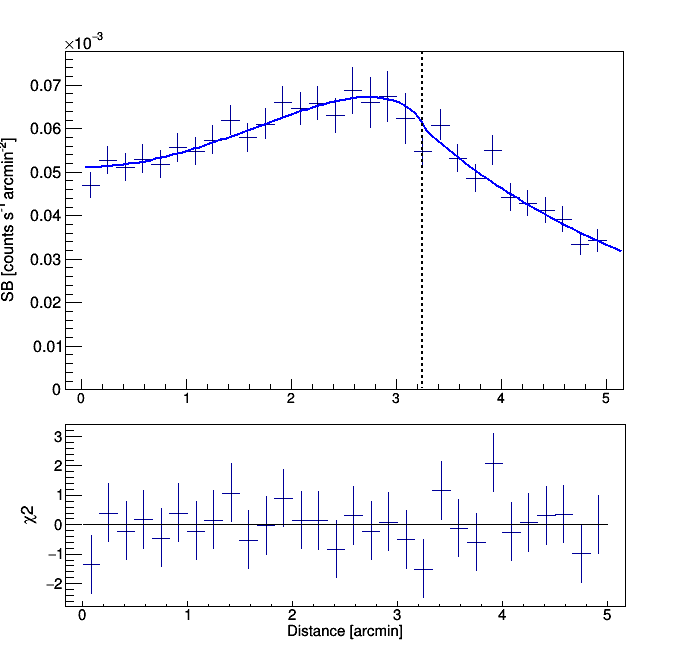}
        \caption{}
        \label{rfidtest_yaxis2}
        \end{subfigure}
    \caption{(a) The box region used for extracting surface brightness profile is shown in green, and the white line shows the location of the shock. Magenta contours correspond to the GMRT radio sources, as mentioned in Fig.~\ref{Xray_radio_img}. (b) (Top) resulting X-ray surface brightness profile as a function of distance and (bottom) corresponding $\chi^{2}$ residuals. The blue line shows the broken power-law best-fit model and the dashed line the location of the shock across the profile.}
    \label{Xray_SB_img}        
\end{figure*}
\end{center}

To understand the effect of the radio lobe on the X-ray bridge, we derived the surface brightness (SB) profile along the extended radio lobe direction (NE) (Fig.~\ref{Xray_SB_img}(a)) and the resulting surface brightness profile (Fig.~\ref{Xray_SB_img}(b)). First, we calculated that the total counts within the bridge region are $\sim$ 13744 and the average background counts are $\sim$ 2690 (in the corresponding blank sky image), giving a SNR\footnote{SNR=Source counts/$\sqrt{\rm Total\; counts}$} of $\sim$ 94. 
We used the PROFFIT v1.5 software \citep{2011A&A...526A..79E} to derive the profile using an image corrected by the exposure map and point source subtracted (to avoid any contamination). 
{We extracted the photon counts from the green box region, as shown in Fig.~\ref{Xray_SB_img}(a). The total number of counts within the box is  $\sim$ 7200. The selection of the box size and angle were taken from the XMM-{\it Newton} 2D temperature map (see Section 5.2), where we noticed a high and low temperature distribution across the north lobe.  The size of the box (w $\times$ h) is 3.8$'$ $\times$ 5$'$ and 170$^{\circ}$ rotation angle with respect to the RA axis}. We took bin size is 10$''$ and then grouped the counts so that each bin has a minimum SNR of 5 and fitted the obtained SB profile with a broken power-law model, as shown in Fig.~\ref{Xray_SB_img}(b). 
 There are five parameters which represent the model: inner slope ($\Gamma1$), outer slope ($\Gamma2$), break radius ($\mathcal{R}_{break}$), surface brightness or density jump ($n_{e,out}$/$n_{e,in}$), and normalisation ($K$). In the fitting process, we kept all parameters free.
\begin{equation}
n_{e}(R) = 
\begin{cases}
 K\eta^{-\Gamma1}, & \text{if }\eta <\mathcal{R}_{break}\\    
 K\eta^{-\Gamma2}, & \text {otherwise}\\
\end{cases}
\end{equation} 

\noindent where $\eta$ is the 3D radius projected on the 2D sky. {The best fitting model values are $\Gamma1$ = -0.92 $\pm$ 0.2, $\Gamma2$ = 0.83 $\pm$ 0.09, $\mathcal{R}_{break}$ = 3.2 $\pm$ 0.32, jump ($n_{e,out}$/$n_{e,in}$) = 1.09 $\pm$ 0.14 and $K$ = -4.46 $\times$ 10$^{-5}$. All these errors are related to fitting procedure errors. The reduced $\chi^{2}$ is 0.64 for 24 degrees of freedom and probability of 90\%.}  This best-fitted surface brightness model indicates that there is a discontinuity present ($\mathcal{R}_{break}$), which is coincident with the radio lobe position. This discontinuity is visible $\sim$ 80-90$''$ (140-160 kpc) away from the outermost contour of the radio lobes. It is located 6$'$ and 5.2$'$ away from the A2384(N) and A2384(S) X-ray peaks, respectively. We confirmed that the location of $\mathcal{R}_{break}$ or discontinuity and jump fit values are do not vary with the increase the box size in the same direction. 

\section{Spectral analysis}

The spectral analysis was restricted to the energy range  [0.7--7.0]\,keV and all the spectra were grouped to contain a sufficient SNR in each energy bin.  In all the fits, the redshift was fixed at $z =$ 0.0943 and the hydrogen column density ($N_{H}$) at the Leiden/Argentine/Bonn (LAB) value ($n_{H}$ = 3.02 $\times$ 10$^{20}$ atoms cm$^{-2}$ ) while the temperature ($kT$), metallicity (Z), and the spectrum normalisation (K) were free to vary.
A single temperature fit was adopted to model the cluster spectrum with an absorbed thermal plasma emission model WABS(APEC).
Abundances were measured assuming the ratios from \citet{Asplund09}. For background subtraction, we used blank sky event files after applying the necessary scaling.

 If we divide the normalisation by the number of pixels in each bin (K/$N_{\rm P}$), we obtain the projected emission measure that is proportional to the square of the electron density $n_{e}^{2}$, integrated along the line of sight. Thus, with the best-fit temperature and normalisation (K), we estimated the pseudo-entropy (S) and pseudo-pressure (P) in the following ways: $S= kT \times [\frac{K}{N_{\rm P}}]^{-1/3}$ and $P = kT \times [\frac{K}{N_{\rm P}}]^{1/2}$.

\subsection{Chandra data}

\par For {\it Chandra} data spectral analysis purposes, we divided the X-ray emission of A2384(N), A2384(S) and bridge into eight regions, as shown in Fig.~\ref{Spectral_plot}(a) and extracted the spectra from each region. We grouped these spectra such that there is a minimum signal to noise ratio (SNR) of 3 in each bin. 

We have plotted kT, S, P, K and abundance (Z) projected radial profiles in Fig.~\ref{Spectral_plot}(b) and given their best-fit values in Table \ref{spectral_val} along with their 68\% confidence interval. We also extracted a single spectrum for the X-ray bridge, over a large region, to study its deprojected thermodynamic properties (Fig.~\ref{Spectral_plot}(a)). We used a box size of ($l$ $\times$ $b$) 290$''$ $\times$ 430$''$ (508 $\times$ 753 kpc$^2$) to extract the spectrum. 


\subsection{XMM-\textit{Newton} 2D maps}
\par We have derived 2D temperature ($kT$), pseudo-entropy ($S$) and pseudo-pressure ($P$) maps for the A2384 cluster (Fig.~\ref{Spectral_2D_plot}). {We have shown temperature error and reduced $\chi^{2}$ maps into Appendix Fig. 1}. This was achieved by dividing the data into small regions from which spectra were extracted \citep[this procedure was already described in][]{Durret10, Durret11, Lagana08, Lagana19}. For XMM-\textit{Newton} data, we excluded the energy band from 1.2 to 1.9 keV to avoid any influence from Al and Si instrumental lines. We set a minimum count number of 900 (to assure at least SNR of 30)
to each pixel and, if necessary, we increased the region up to a box of 5 $\times$ 5 pixels. Each pixel has a size of 12.8 $\times$ 12.8 arcsec$^2$, corresponding to 22.4 $\times$ 22.4 kpc$^2$.
If we still do not have enough counts after binning, the pixel is ignored and we proceed to the next neighbouring pixel. When we have enough counts, the spectra of  MOS1, MOS2 and pn are then simultaneously fitted, and the best $kT$ and $Z$ values are attributed to the central pixel. 

\section{Discussion}

\subsection{Dynamical state of A2384}
\par
\begin{center}
\begin{figure*}
    \centering
    \begin{subfigure}[t]{0.50\textwidth}
        \includegraphics[width=1\textwidth]{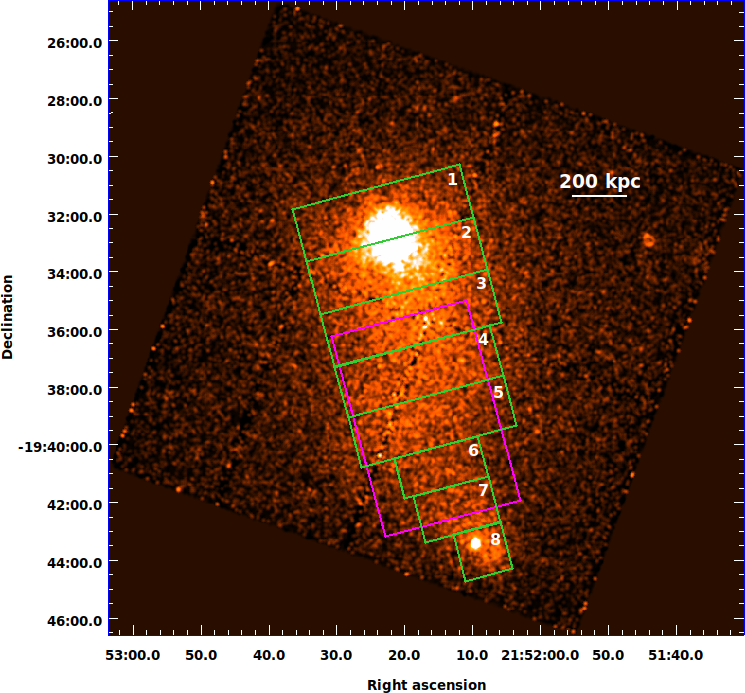}
        \caption{}
        \label{rfidtest_xaxis1}
    \end{subfigure}
    \begin{subfigure}[t]{0.40\textwidth}
        \includegraphics[width=1\textwidth]{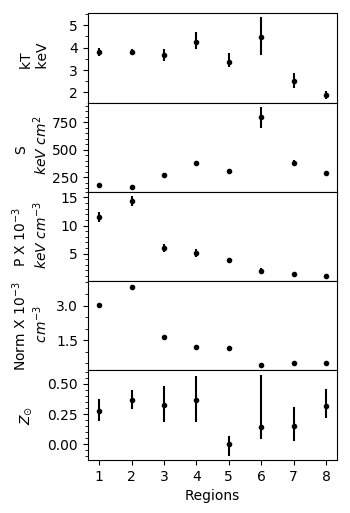}
        \caption{}
        \label{rfidtest_yaxis2}
        \end{subfigure}
    \begin{subfigure}[t]{0.60\textwidth}
        \includegraphics[width=1\textwidth]{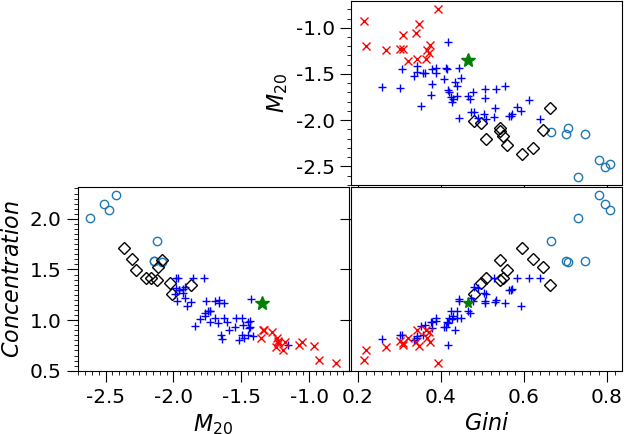}
        \caption{}
        \label{rfidtest_yaxis2}
        \end{subfigure}
    \caption{(a) Regions used to extract the spectra from {\it Chandra} data. The green boxes (marked 1 to 8) and the magenta box were considered to obtain projected and de-projected spectral values. (b) Projected spectral profiles for regions 1 to 8. (c) {Three morphology parameters plotted in the parameter-parameter planes. $\circ$ = strongly relaxed  (strong cool core) clusters;  {\small $\lozenge$} = relaxed (cool core) clusters;  {\small \color {blue} $+$} = non-relaxed (weak cool core) clusters; and {\small \color {red} $\times$} = strongly non-relaxed (non-cool core) clusters. A2384(N) is marked with {\color{green} $\bigstar$}. See text for details.}}
    \label{Spectral_plot}        
\end{figure*}
\end{center}

 A2384 is a very peculiar, bimodal cluster that is elongated along the north to south direction with a size of $\sim$~1.5 ~Mpc. A2384(N) is the most massive cluster of the system. Analysis of the morphology of A2384 can lead to an understanding of its merging stage, evolution and degree of disturbance. To do so, we performed a joint analysis using radio and X-ray data to better understand the radio lobe interaction with the X-ray bridge of A2384.
 We used three (Gini, $M_{20}$ and Concentration) morphology parameters \citep[and references therein]{2015A&A...575A.127P} to characterise the dynamical state of A2384(N) and to compare its position with the morphologies of other clusters in the parameter planes. These parameters trace the flux distribution for a given region and their values depend on the degree of disturbance in a given cluster image. We refer the reader to \cite{2015A&A...575A.127P} for more details about these parameters and their detailed calculation. To calculate these parameters, we considered a 500 kpc region around the centroid of A2384(N). For the purpose of comparing the morphology of A2384 with other clusters, we used data from \cite{2015A&A...575A.127P} and categorised the dynamical state of each cluster into four stages: strongly relaxed cluster, relaxed cluster, non-relaxed cluster, and strongly non-relaxed cluster, according to the combination of these three parameters. As seen in Fig.~\ref{Spectral_plot}(c), strongly relaxed clusters are well separated from strongly non-relaxed clusters. A2384(N) falls in the non-relaxed cluster category. The three parameter values (with their uncertainties) are listed in Table \ref{A2384_X-ray_morph}. Based on the cluster dynamical state separation definitions described in \cite{2015A&A...575A.127P}, the three morphology parameters indicate that A2384(N) is a non-relaxed disturbed cluster. {Further, in this analysis, the value of the $M_{20}$ parameter is high as compared with the Gini and Concentration parameters for the same dynamical class of cluster. The Gini and Concentration parameters indicate that A2384 is a non-relaxed cluster (based on its visualisation also it looks like a weak merger), but the higher value of $M_{20}$ suggests that it is strongly non-relaxed. $M_{20}$ is defined as the normalised second-order moment of the brightest 20\% of the total cluster flux. In the case of the bimodal cluster A2384, the centroid is located towards the bridge and is $\sim$ 1.2$'$ away from the X-ray peak emission (where 20\% of the brightest flux is within only a few pixels). This large distance between the centroid and X-ray peak (due to the bimodal nature of A2384) leads to a higher value of $M_{20}$. }
 This morphology results could support the scenario that A2384 is a post-merger cluster. The low mass cluster A2384(S) has passed through the more massive cluster A2384(N) and probably stripped it of a large amount of hot gas (and galaxies) along the merger direction. 

\par Further, typically the offset of the BCG from the X-ray peak is consistent with a merging cluster \citep{2012MNRAS.420.2120M,2016MNRAS.457.4515R}. The A2384(N) BCG 
(LEDA 190740) is 10-15$''$ away from the X-ray peak, as also noticed by \cite{2011A&A...525A..79M}. We also noticed that the position of A2384(S) BCG (LEDA 67523) is 40$''$ far from its X-ray peak. These authors also derived the same conclusion that A2384 is a post-merger cluster seen $\sim$ 1-1.5 Gyr after core collapse. They have also noticed that (1) the velocity dispersion of both subclusters - A2384(N) and (S) - is very high ($\sim$ 1000 km s$^{-1}$), and (2) the radial velocity of the BCG of A2384(N) is offset (by $\sim$ 600 km s$^{-1}$) from the mean velocity of the member galaxies. These two properties indicate that A2384 is a disturbed system and that the merging has already happened.

\begin{center}
\begin{figure*}
    \centering
    \begin{subfigure}[t]{0.45\textwidth}
        \includegraphics[width=1\textwidth]{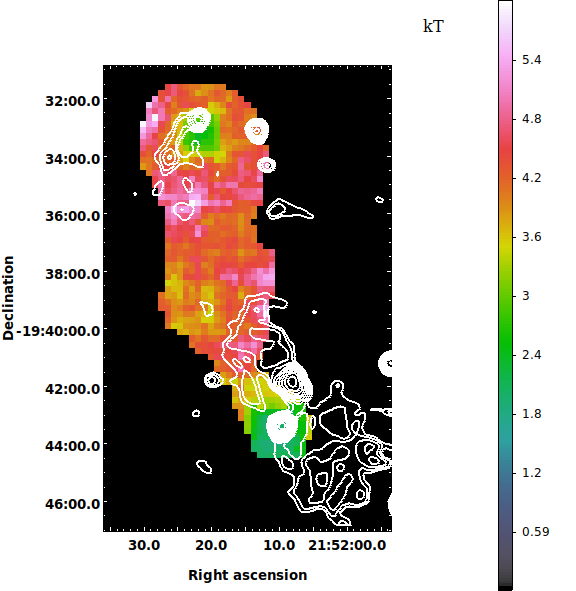}
        \caption{}
        \label{rfidtest_xaxis1}
    \end{subfigure}
    \begin{subfigure}[t]{0.45\textwidth}
        \includegraphics[width=1\textwidth]{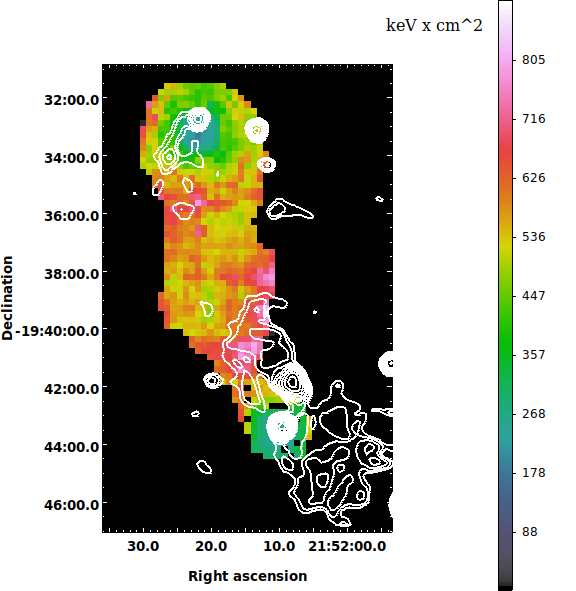}
        \caption{}
        \label{rfidtest_yaxis2}
        \end{subfigure}
    \begin{subfigure}[t]{0.45\textwidth}
        \includegraphics[width=1\textwidth]{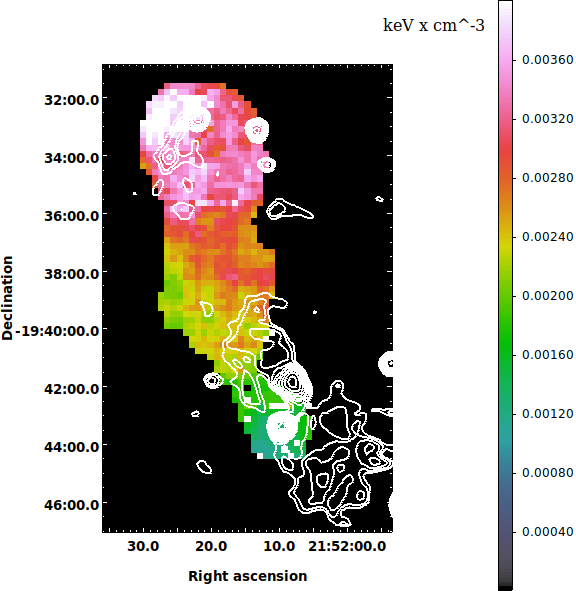}
        \caption{}
        \label{rfidtest_yaxis2}
        \end{subfigure}
    \begin{subfigure}[t]{0.45\textwidth}
        \includegraphics[width=1\textwidth]{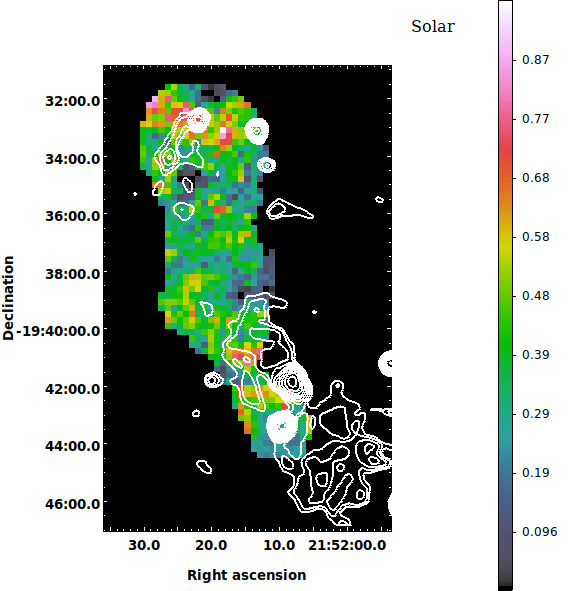}
        \caption{}
        \label{rfidtest_yaxis2}
        \end{subfigure}

    \caption{Thermodynamic maps derived from XMM-{\it Newton} data.(a) Temperature map, (b) entropy (S) map, (c) pressure (P) map and (d) metallicity (Z) map. The white contours are those of the 325 MHz radio image as in Fig.~\ref{Xray_radio_img}. }
    \label{Spectral_2D_plot}        
\end{figure*}
\end{center}


\begin{table}
\centering
\caption{A2384 morphology parameter values.}
\begin{center}
\begin{tabular}{@{}cccccccc@{}}
\hline
\hline 
Gini &$M_{20}$ & Concentration \\
\hline
0.46$\pm$0.0035 & -1.34$\pm$0.027 & 1.17$\pm$0.009\\
\hline 
\end{tabular}
\end{center}
\label{A2384_X-ray_morph} 
\end{table}


\subsection{X-ray properties of the bridge}
\par Many clusters show a discontinuity in their X-ray surface brightness profiles, which often indicates the presence of shocks and/or cold fronts \citep{2001ApJ...563...95M,2002AstL...28..495V,2007PhR...443....1M}. The density jump ($n_{\rm e,out}$/$n_{\rm e,in}$) is considered to be due to the propagation of a shock through the dense gas. We also see a discontinuity in the surface brightness profile across the bridge of A2384, as shown in Fig.~\ref{Xray_SB_img}. This is similar to other merging clusters reported in the literature.  In Fig.~\ref{Xray_SB_img}(a), we marked the position of the shock where we observed a discontinuity in the corresponding surface brightness profile (Fig.~\ref{Xray_SB_img}(b)) at a distance of 3.2$'$ from the origin of the sector. We derived the Mach number based on the compression factor (or density jump) as:
\begin{equation}
{\mathcal{M} = \bigg(\frac{2C}{\gamma+1-(\gamma-1)C}\bigg)^{1/2}}
\end{equation}  
This gives $\mathcal{M}$ = 1.06 $\pm$ 0.09 with the assumption of an adiabatic index for fully ionised
plasma of $\gamma$ = 5/3 and C is the compression factor. Typically, a powerful merger shock produces a shock wave with $\mathcal{M}$ in the range 2-3 \citep{2019SSRv..215...16V}. However, in the case of the A2384 X-ray bridge, we found a low Mach number. This could be due to a weak shock propagated by the radio lobe of the FR I galaxy. \cite{2004rcfg.proc..181B} has reported a weak shock with $\mathcal{M}$ of $\sim$ 1.27 in the cool core Hydra A cluster. Recently \cite{2018A&A...609A..61C,2019A&A...622A..25W} also reported weak shock propagation in several clusters.

\par {To study the thermodynamic properties of the bridge, we considered a total of 8 regions along the A2384 north to south extended area (Fig.~\ref{Spectral_plot}(a)) and examined the spectral properties within each of the zones using {\it Chandra} data. The temperature varies between $\sim$ 3 and 4 keV from regions 1 to 6. Regions 1 and 2 correspond to the A2384(N). Region~3 is the overlap between the A2384(N) and bridge. The average temperature of A2384(N) is $\sim$ 3.8 keV. The temperature is higher in region 4, which is a part of the bridge. Regions 5 and 6 correspond to the {radio-lobe/ICM interaction site} where the temperature is $\sim$ 4.47 keV in region 6 and  drops in region 5 ($\sim$ 3.3 keV). Region 6, which has the highest temperature, is the narrowest region of the bridge. Regions 7 and 8 correspond to A2384(S) and have an average temperature of $\sim$ 2.1 keV. \cite{2011A&A...525A..79M} estimated the temperatures (based on XMM-{\it Newton} data) of A2384(N) and (S) to be $\sim$ 3-4 keV, the surrounding gas residing in the bridge being hotter, with a temperature up to 4-5 keV. This previous finding is similar to our temperature estimation.}

The normalisation parameter and pressure both decrease in the bridge regions. However, there is an indication that the entropy increases in the bridge regions and maximum at region 6. This indicates that the FR I radio lobe has supplied energy to the hot X-ray plasma. This adiabatic non-thermal heating could deposit energy into the hot gas, causing the entropy to increase in the interaction region. We also plotted the metallicity distribution in A2384, and found a solar abundance of $\sim$ 0.1-0.2 $Z_{\odot}$ in the bridge region. However, due to poor counts, there is a large uncertainties with metalicity estimations.
\par Apart from the projected radial profile, we also derived the de-projected thermodynamic properties of the bridge region: the temperature $\sim$ 4.11$^{0.17}_{-0.17}$ keV, abundance $\sim$ 0.22$^{0.09}_{-0.08}$ $Z_{\odot}$, entropy S $\sim$ 353$^{11}_{-11}$ $keV$ cm$^{2}$, pressure P $\sim$ 4.82$^{0.272}_{-0.266}$ $\times$ 10$^{-3}$ $keV$ $cm^{-3}$, electron density $n_{e}$ $\sim$ 1.17$^{0.016}_{-0.016}$ $\times$ 10$^{-3}$ cm$^{-3}$ and integrated mass $\sim$ 0.56$^{0.008}_{-0.008}$ $\times$ 10$^{13}$ $M_{\odot}$. We also estimated the luminosity of the bridge to be $\sim$ 9.22 $\times$ 10$^{43}$ erg s$^{-1}$. To calculate these values, we assumed the volume of the filament to be a cylinder ($\pi$r$^{2}$h) of size 145$''$(r) $\times$ 430$''$(h) (254 $\times$ 753 kpc$^3$). 
Then we calculated the electron density as:
\begin{equation}
n_{e} = \sqrt{\frac{Norm \times 4\pi \times D_{A}^{2} \times (1+z)^{2} \times 10^{14}}{Volume \times 1.18}}
\end{equation}

\noindent where $D_{A}$ is the angular diameter distance to the source (cm), Norm is the normalisation factor and 1.18 is the electron to proton ratio ($n_{e}/n_{p}$). The integrated total mass within the bridge is calculated as $m_{N}n_{e}V$ where $m_{N}$ is the average nucleon mass per hydrogen atom (2.12 $\times$ 10$^{-24}$ g/H atom), $n_{e}$ the electron density (cm$^{-3}$) and $V$ the volume of the cylinder (cm$^3$).

\par We generated 2D thermodynamic maps of A2384 based on XMM-{\it Newton} data as shown in Fig.~\ref{Spectral_2D_plot}. There is a arc of the size $\sim$ 3.8$'$ visible in the temperature and entropy maps where temperature and entropy are high. This arc well overlaps with the north radio lobe of FR I galaxy. There is currently no evidence that the FR I radio source is strongly shocking the ICM, but the possibility of weaker shocks cannot be ruled out. If we assume a weak shock is propagating in the direction of the radio lobe extension (north-east), then we can estimate the mean temperature within the lobe region, {$T_{post}$ $\sim$ 4.4 $\pm$ 0.5 keV and outside of the lobe region, $T_{pre}$ $\sim$ 3.5 $\pm$ 0.3 keV}. These values are similar to the temperature values derived from the {\it Chandra} data (Regions 5 and 6 in Fig.~\ref{Spectral_plot}(a)). The standard Rankine-Hugoniot model predicts the temperature jump condition:
\begin{equation}
\frac{T_{post}}{T_{pre}} = \frac{[2\gamma\mathcal{M}^2-(\gamma-1)][(\gamma-1)\mathcal{M}^2+2]}{(\gamma+1)\mathcal{M}^2}
\end{equation}

This leads to a threshold on the Mach number $\mathcal{M}$ $\sim$ 1.25 $\pm$ 0.035 with the assumption of an adiabatic index of $\gamma$ = 5/3. This Mach number value is close to that derived from the surface brightness profile of the {\it Chandra} image (within the error bar). There is also good agreement between the position of the shock in the {\it Chandra} emissivity map and the {\it XMM}-Newton temperature map.  The pressure map also shows high values around the radio lobe (towards the north direction), which could be due to the impelling of the ICM by the radio lobe. In the XMM-{\it Newton} metallicity map, the solar abundance values in the bridge region are higher than those derived from {\it Chandra} data. 

Accretion shocks heat the cluster outskirts {($>$
$R_{200}$) \citep{2009ApJ...696.1640M}} raising entropies above 1000 keV cm$^{2}$. This entropy is expected to increase further out to the virial radius. In the A2384
case, the overall low entropy of the bridge suggests that the radiating hot gas in the A2384 bridge has not yet been reached by the shock-heating operating on the ICM in the cluster outskirts, but there could be a localised entropy enhancement due to the radio lobe interaction. 

\subsection{Masses of the A2384 substructures}
{\cite{2014A&A...570A..40P} and \cite{2011A&A...525A..79M} tried to estimate the masses of A2384 (total) and its subclusters. These mass calculations are based on velocity dispersions. A2384 is a disturbed system, so the velocity distribution of this system is largely affected by the cluster merging effect. Observations and simulations both have shown that during the cluster merging process the kinematics of galaxies of an entire cluster system can be strongly boosted (by a factor of $\sim$ 2) and large scale gas motion energy (of colliding clusters) is converted into random motion of the galaxies \citep[and reference therein]{2019SSRv..215...25P}. This results in a significant mass overestimation for a merging system. \cite{2014A&A...570A..40P} and \cite{2011A&A...525A..79M} have also noted this problem in their mass calculations, due to the overestimation of the velocity dispersion in A2384. Hence, the mass of A2384 is overestimated and the mass ratio of the A2384(N) and A2384(S) subclusters is an upper limit. 

 \par The mass of A2384, as reported in the MCXC catalogue \citep{2011A&A...534A.109P}, is $M_{500}$ $\sim$ 2.61 $\times$ 10$^{14}$ $M_{\odot}$. This is the total mass (within $r_{500}$\footnote{the characteristic radius within which the mean over-density of the cluster is 500 times the critical density at the cluster redshift}) of the cluster (derived from the Mass-Luminosity scaling relation) which includes dark matter, X-ray gas and stellar mass. To calculate the gas mass, we used the \cite{2009ApJ...692.1033V} sample and the $M_{500}$-$M_{gas}$ scaling relation given in Table 3 of that paper\footnote{$f_{g} = M_{gas}/M_{500} = [(0.0764 \pm 0.004){\it h}^{-1.5} + (0.037 \pm 0.006) \times log(M_{500}/10^{15}$)]}. From this, we estimated that the gas mass of the entire A2384 system is $M_{gas,total}$ $\sim$ 2.84$\pm 0.09$ $\times$ 10$^{13}$ $M_{\odot}$.  After subtracting the gas mass of the bridge computed in the previous section, $M_{gas,bridge}$,  from the total gas mass of A2384, we find $M_{gas,N+S}$ $\sim$ 2.28$\pm 0.08$ $\times$ 10$^{13}$ $M_{\odot}$, which includes only the gas masses of A2384(N) and (S). This implies that the gas mass of the X-ray bridge is $\sim$ 4 times lower than that of A2384 (N+S).
 
\cite{2014A&A...570A..40P} reported that the total masses of A2384(N) and A2384(S) are $M_{200}$ $\sim$ 1.84 $\times$ 10$^{15}$ $M_{\odot}$ and $\sim$ 1.17 $\times$ 10$^{15}$ $M_{\odot}$, respectively. Hence, their mass ratio is 1.6:1. If we assume the same mass ratio for the X-ray gas, then $M_{gas}$ of A2384(N) is $\sim$ 1.42$\pm 0.05$ $\times$ 10$^{13}$ $M_{\odot}$, that is $\sim$ 2.5 times higher than the mass of the X-ray bridge. The gas mass of A2384(S) is $\sim$ 0.86$^{0.02}_{-0.03}$ $\times$ 10$^{13}$ $M_{\odot}$, that is $\sim$ 1.5 times higher than the mass of the X-ray bridge. From the MCXC catalogue, the total luminosity ($L_{500}$) of the A2384 system is 1.66 $\times$ 10$^{44}$ erg s$^{-1}$. Hence, the luminosity of the bridge is roughly 1.8 times lower than that of the entire A2384 system. }

\subsection{Impact of the radio lobe on the X-ray bridge}
\par Combined radio, X-ray and optical observations often show that the fast motion of the (active) radio galaxies and the interaction of the radio jet outflows with the ICM result in an asymmetry in the alignment of the radio lobes. The dense ICM can change the morphology and physical nature of the radio lobes via exerting ram pressure.
In the case of A2384 the FR I radio galaxy LEDA 851827 pushes the ICM in the east direction through one of its lobes, as opposed to the jet bending scenario where the ICM distorts the radio jets. Here the radio lobe exerts a mechanical work on the X-ray bridge and displaces the hot and dense gas, and finally, it could be mixed with the surrounding medium and deposit cosmic rays and magnetic energy in the ICM \citep{2012NJPh...14e5023M}. 

{ The radial velocity of LEDA 851827 is 28,321 $\pm$ 18 km s$^{-1}$ \citep{2014A&A...570A..40P}, which is close to the radial velocity (28,696 $\pm$ 53 km s$^{-1}$) of the BCG of A2384(S), and seems to be associated with the A2384(S) cluster. From NVSS (1400 MHz), we estimate the flux density of the radio galaxy core to be 30 $\pm$ 0.34~mJy. In the 325 MHz GMRT observation, the flux density is 116 $\pm$ 7.3 mJy. This gives a spectral index $\alpha$ ($S_{\nu}$ $\propto$ $\nu^{\alpha}$) $\alpha_{325MHz}^{1400MHz}$ $\sim$ -0.92 for the host radio galaxy. We used CASA's IMFIT task to estimate the integrated flux density of the radio galaxy core. Its position is slightly away from the X-ray emission (100$''$), and north of the BCG of A2384(S). We estimated the radio luminosity as P = 4$\pi$ $D_{L}^{2}$ $S_{\nu}$ where $D_{L}$ is the luminosity distance at the cluster redshift and $S_{\nu}$ the total integrated flux density of the source at the given frequency $\nu$. We measured the total radio luminosity of the north lobe (which interacts with the ICM) to be $P_{325MHZ}$ $\sim$ 2.5 $\times$ 10$^{28}$ erg s$^{-1}$ Hz$^{-1}$. We extrapolated this to 1400 MHz with a spectral index of $-0.92$, assuming the same spectral index as the radio galaxy core, and obtained $P_{1400MHz}$ $\sim$ 6.5 $\times$ 10$^{27}$ erg sec$^{-1}$ Hz$^{-1}$. We estimated the radio luminosity of the south lobe to be $P_{325MHZ}$ $\sim$ 3.8 $\times$ 10$^{28}$ erg sec$^{-1}$ Hz$^{-1}$ and $P_{1400MHZ}$ $\sim$ 1.0 $\times$ 10$^{28}$ erg sec$^{-1}$ Hz$^{-1}$. The south lobe is $\sim$ 1.5 times more luminous than and roughly twice as large as the north lobe. This could suggest that the size of the north lobe is smaller because of its interaction with the ICM and the lobe cannot expand due to the opposite pressure exerted on it by the surrounding dense ICM.} 
\par More radio observations of cluster filaments are needed to understand how the shocks (internal and external) govern the large scale structure formation process \citep{2000ApJ...542..608M,2003ApJ...593..599R,2006MNRAS.367..113P}. It is essential to investigate the presence of non-thermal structures beyond the Mpc scale of clusters and to link them with lower density regions. \cite{2010ApJ...724L.143E} reported the first discovery of a bent double lobe radio source (DLRS) between A1763 and A1770 clusters' filament. They used properties of the radio source to constrain the density of the surrounding interfilament medium. The extended radio lobe which interacts with the X-ray bridge or filament between A2384(N) and A2384(S) is a unique detection, and it can be used to set the limits on the pressure of the hot gas, shock structures and large scale magnetic fields.

\subsection{Future radio observations}
\par It is important to derive the dynamical age of the interacting radio galaxy of A2384 to understand its evolution with respect to the ICM, the active cycles of the galactic nuclei, and the physical state of the surrounding galactic and intergalactic medium, which allows the launch and propagation of jets. In a follow-up analysis of the low frequency wideband uGMRT data, we plan to model the radio lobes. We also need high-frequency wideband radio data to study the shape of the spectrum along the radio lobes and to conduct the classical spectral-ageing analysis \citep{2016MNRAS.458.4443H,2017MNRAS.469..639H}.

\section{Conclusions}
\par In this paper, we studied the X-ray properties of the A2384 bridge - a high density region extended between the clusters A2384(N) and A2384(S). We have also shown GMRT radio observations of this system and found radio-X-ray interaction at the cluster outskirts. The north lobe of the FR I radio galaxy displaces the hot gas on a large scale in the bridge region. X-ray cavities are the classic example of radio-X-ray interaction, but in this rare case, this interaction happens in the cluster outskirts region. Further, unlike bent-tail radio galaxies, the FR I radio galaxy causes a displacement in the X-ray gas bridge region. Based on thermodynamic 1D profiles ({\it Chandra}) and 2D maps (XMM-{\it Newton}), the entropy increases at the {the radio-lobe/ICM interaction site}, which indicates the transportation of non-thermal energy to the thermal plasma. Furthermore, the high surrounding pressure suggests a displacement of hot gas. We also derived Mach numbers based on the surface brightness profile and temperature maps, which suggest the presence of a weak shock due to the radio lobe interaction. To better understand this system, we need more radio data at multiple frequencies.


\begin{table*}
\caption{Spectral properties of the A2384.}
\begin{tabular}{ccccccccc}
\hline
\hline
Region & T &Normalisation &$Z_{\odot}$& $S$& $P$ & $\chi^{2}$/dof\\
& keV &  $\times$10$^{-3}$ $cm^{-3}$ & &keV $cm^{2}$ &keV $cm^{-3}$ & $\times$10$^{-3}$&  \\
\hline
1 & 3.81$^{0.18}_{-0.18}$ &   3.030$^{0.092}_{-0.091}$ & 0.28$^{0.09}_{-0.09}$ &  182$^{4.83}_{-5.05}$ &   11.556$^{0.915}_{-0.877}$ & 119/190 \\
2 & 3.80$^{0.14}_{-0.14}$ &   3.783$^{0.094}_{-0.094}$ & 0.37$^{0.08}_{-0.08}$ &  156$^{3.28}_{-3.37}$ &   14.372$^{0.914}_{-0.887}$ & 131/212 \\
3 & 3.67$^{0.28}_{-0.26}$ &   1.629$^{0.082}_{-0.082}$ & 0.32$^{0.16}_{-0.14}$ &  265$^{10.64}_{-10.47}$ & 5.983$^{0.774}_{-0.709}$ & 98/147 \\
4 & 4.26$^{0.41}_{-0.32}$ &   1.205$^{0.064}_{-0.063}$ & 0.36$^{0.20}_{-0.18}$ &  376$^{22.46}_{-15.42}$ & 5.138$^{0.799}_{-0.630}$ & 91/128 \\
5 & 3.36$^{0.37}_{-0.23}$ &   1.145$^{0.042}_{-0.040}$ & 0.00$^{0.06}_{-0.10}$ &  307$^{25.98}_{-14.24}$ & 3.853$^{0.582}_{-0.388}$ & 65/109 \\
6 & 4.47$^{0.88}_{-0.79}$ &   0.417$^{0.040}_{-0.044}$ & 0.14$^{0.44}_{-0.10}$ &  800$^{101.33}_{-90.09}$ &1.865$^{0.578}_{-0.489}$ & 33/50 \\
7 & 2.50$^{0. 35}_{-0.29}$ &   0.530$^{0.051}_{-0.049}$ & 0.15$^{0.16}_{-0.12}$ &  381$^{27.56}_{-21.82}$ & 1.323$^{0.329}_{-0.262}$ & 29/51 \\
8 & 1.88$^{0.15}_{-0.16}$ &   0.539$^{0.053}_{-0.051}$ & 0.32$^{0.14}_{-0.10}$ &  284$^{4.63}_{-6.98}$ &   1.014$^{0.191}_{-0.176}$ & 32/45 \\
\hline
\end{tabular}
\label{spectral_val}
\end{table*}

\section*{Acknowledgments}
VP and KT acknowledge the SKA SA Fellowship. TFL acknowledges
financial support from the Brazilian agencies FAPESP and CNPq through grants 2018/02626-8 and 303278/2015-3, respectively. FD is grateful to CNES for long-term support. We would like to thank the anonymous referee for his/her carefully reading our manuscript and for giving such constructive comments which substantially helped improving the quality of the paper.
This research has made use of the data from {\it Chandra} Archive and of software provided by the {\it Chandra} X-ray Centre (CXC) in the application packages CIAO, ChIPS, and Sherpa. Based on observations obtained with XMM-{\it Newton}, an ESA science mission with instruments and contributions directly funded by ESA Member States and NASA. We thank the staff of the GMRT who have made these observations possible. GMRT is run by the National Centre for Radio Astrophysics of the Tata Institute of Fundamental Research.

\bibliography{references}

\begin{thebibliography}{}
\makeatletter
\relax
\def\mn@urlcharsother{\let\do\@makeother \do\$\do\&\do\#\do\^\do\_\do\%\do\~}
\def\mn@doi{\begingroup\mn@urlcharsother \@ifnextchar [ {\mn@doi@}
  {\mn@doi@[]}}
\def\mn@doi@[#1]#2{\def\@tempa{#1}\ifx\@tempa\@empty \href
  {http://dx.doi.org/#2} {doi:#2}\else \href {http://dx.doi.org/#2} {#1}\fi
  \endgroup}
\def\mn@eprint#1#2{\mn@eprint@#1:#2::\@nil}
\def\mn@eprint@arXiv#1{\href {http://arxiv.org/abs/#1} {{\tt arXiv:#1}}}
\def\mn@eprint@dblp#1{\href {http://dblp.uni-trier.de/rec/bibtex/#1.xml}
  {dblp:#1}}
\def\mn@eprint@#1:#2:#3:#4\@nil{\def\@tempa {#1}\def\@tempb {#2}\def\@tempc
  {#3}\ifx \@tempc \@empty \let \@tempc \@tempb \let \@tempb \@tempa \fi \ifx
  \@tempb \@empty \def\@tempb {arXiv}\fi \@ifundefined
  {mn@eprint@\@tempb}{\@tempb:\@tempc}{\expandafter \expandafter \csname
  mn@eprint@\@tempb\endcsname \expandafter{\@tempc}}}

\bibitem[\protect\citeauthoryear{{Akahori} \& {Yoshikawa}}{{Akahori} \&
  {Yoshikawa}}{2010}]{2010PASJ...62..335A}
{Akahori} T.,  {Yoshikawa} K.,  2010, \mn@doi [\pasj] {10.1093/pasj/62.2.335},
  \href {http://adsabs.harvard.edu/abs/2010PASJ...62..335A} {62, 335}

\bibitem[\protect\citeauthoryear{{Akamatsu} et~al.,}{{Akamatsu}
  et~al.}{2016}]{2016A&A...593L...7A}
{Akamatsu} H.,  et~al., 2016, \mn@doi [\aap] {10.1051/0004-6361/201629275},
  \href {http://adsabs.harvard.edu/abs/2016A%26A...593L...7A} {593, L7}

\bibitem[\protect\citeauthoryear{{Asplund}, {Grevesse}, {Sauval}  \&
  {Scott}}{{Asplund} et~al.}{2009}]{Asplund09}
{Asplund} M.,  {Grevesse} N.,  {Sauval} A.~J.,   {Scott} P.,  2009, \mn@doi
  [\araa] {10.1146/annurev.astro.46.060407.145222}, \href
  {http://adsabs.harvard.edu/abs/2009ARA%26A..47..481A} {47, 481}

\bibitem[\protect\citeauthoryear{{Bagchi}, {Sankhyayan}, {Sarkar},
  {Raychaudhury}, {Jacob}  \& {Dabhade}}{{Bagchi}
  et~al.}{2017}]{2017ApJ...844...25B}
{Bagchi} J.,  {Sankhyayan} S.,  {Sarkar} P.,  {Raychaudhury} S.,  {Jacob} J.,
  {Dabhade} P.,  2017, \mn@doi [\apj] {10.3847/1538-4357/aa7949}, \href
  {http://adsabs.harvard.edu/abs/2017ApJ...844...25B} {844, 25}

\bibitem[\protect\citeauthoryear{{Blanton}}{{Blanton}}{2004}]{2004rcfg.proc..181B}
{Blanton} E.~L.,  2004, in {Reiprich} T.,  {Kempner} J.,   {Soker} N.,  eds,
  The Riddle of Cooling Flows in Galaxies and Clusters of galaxies.
  (\mn@eprint {} {astro-ph/0402342})

\bibitem[\protect\citeauthoryear{{Blanton}, {Gregg}, {Helfand}, {Becker}  \&
  {White}}{{Blanton} et~al.}{2003}]{2003AJ....125.1635B}
{Blanton} E.~L.,  {Gregg} M.~D.,  {Helfand} D.~J.,  {Becker} R.~H.,   {White}
  R.~L.,  2003, \mn@doi [\aj] {10.1086/368140}, \href
  {http://adsabs.harvard.edu/abs/2003AJ....125.1635B} {125, 1635}

\bibitem[\protect\citeauthoryear{{Condon}, {Cotton}, {Greisen}, {Yin},
  {Perley}, {Taylor}  \& {Broderick}}{{Condon}
  et~al.}{1998}]{1998AJ....115.1693C}
{Condon} J.~J.,  {Cotton} W.~D.,  {Greisen} E.~W.,  {Yin} Q.~F.,  {Perley}
  R.~A.,  {Taylor} G.~B.,   {Broderick} J.~J.,  1998, \mn@doi [\aj]
  {10.1086/300337}, \href {http://adsabs.harvard.edu/abs/1998AJ....115.1693C}
  {115, 1693}

\bibitem[\protect\citeauthoryear{{Cuciti}, {Brunetti}, {van Weeren},
  {Bonafede}, {Dallacasa}, {Cassano}, {Venturi}  \& {Kale}}{{Cuciti}
  et~al.}{2018}]{2018A&A...609A..61C}
{Cuciti} V.,  {Brunetti} G.,  {van Weeren} R.,  {Bonafede} A.,  {Dallacasa} D.,
   {Cassano} R.,  {Venturi} T.,   {Kale} R.,  2018, \mn@doi [\aap]
  {10.1051/0004-6361/201731174}, \href
  {https://ui.adsabs.harvard.edu/abs/2018A&A...609A..61C} {609, A61}

\bibitem[\protect\citeauthoryear{{Cypriano}, {Sodr{\'e}}, {Kneib}  \&
  {Campusano}}{{Cypriano} et~al.}{2004}]{2004ApJ...613...95C}
{Cypriano} E.~S.,  {Sodr{\'e}} Jr. L.,  {Kneib} J.-P.,   {Campusano} L.~E.,
  2004, \mn@doi [\apj] {10.1086/422896}, \href
  {http://cdsads.u-strasbg.fr/abs/2004ApJ...613...95C} {613, 95}

\bibitem[\protect\citeauthoryear{{Dehghan}, {Johnston-Hollitt}, {Franzen},
  {Norris}  \& {Miller}}{{Dehghan} et~al.}{2014}]{2014AJ....148...75D}
{Dehghan} S.,  {Johnston-Hollitt} M.,  {Franzen} T.~M.~O.,  {Norris} R.~P.,
  {Miller} N.~A.,  2014, \mn@doi [\aj] {10.1088/0004-6256/148/5/75}, \href
  {http://adsabs.harvard.edu/abs/2014AJ....148...75D} {148, 75}

\bibitem[\protect\citeauthoryear{{Dolag}, {Bykov}  \& {Diaferio}}{{Dolag}
  et~al.}{2008}]{2008SSRv..134..311D}
{Dolag} K.,  {Bykov} A.~M.,   {Diaferio} A.,  2008, \mn@doi [\ssr]
  {10.1007/s11214-008-9319-2}, \href
  {http://adsabs.harvard.edu/abs/2008SSRv..134..311D} {134, 311}

\bibitem[\protect\citeauthoryear{{Durret}, {Lagan{\'a}}, {Adami}  \&
  {Bertin}}{{Durret} et~al.}{2010}]{Durret10}
{Durret} F.,  {Lagan{\'a}} T.~F.,  {Adami} C.,   {Bertin} E.,  2010, \mn@doi
  [\aap] {10.1051/0004-6361/201014566}, \href
  {http://adsabs.harvard.edu/abs/2010A%26A...517A..94D} {517, A94}

\bibitem[\protect\citeauthoryear{{Durret}, {Lagan{\'a}}  \& {Haider}}{{Durret}
  et~al.}{2011}]{Durret11}
{Durret} F.,  {Lagan{\'a}} T.~F.,   {Haider} M.,  2011, \mn@doi [\aap]
  {10.1051/0004-6361/201015978}, \href
  {http://adsabs.harvard.edu/abs/2011A%26A...529A..38D} {529, A38}

\bibitem[\protect\citeauthoryear{{Eckert}, {Molendi}  \& {Paltani}}{{Eckert}
  et~al.}{2011}]{2011A&A...526A..79E}
{Eckert} D.,  {Molendi} S.,   {Paltani} S.,  2011, \mn@doi [\aap]
  {10.1051/0004-6361/201015856}, \href
  {https://ui.adsabs.harvard.edu/abs/2011A%26A...526A..79E} {526, A79}

\bibitem[\protect\citeauthoryear{{Edwards}, {Fadda}  \& {Frayer}}{{Edwards}
  et~al.}{2010}]{2010ApJ...724L.143E}
{Edwards} L.~O.~V.,  {Fadda} D.,   {Frayer} D.~T.,  2010, \mn@doi [\apjl]
  {10.1088/2041-8205/724/2/L143}, \href
  {http://adsabs.harvard.edu/abs/2010ApJ...724L.143E} {724, L143}

\bibitem[\protect\citeauthoryear{{Fabian}}{{Fabian}}{1994}]{1994ARA&A..32..277F}
{Fabian} A.~C.,  1994, \mn@doi [\araa] {10.1146/annurev.aa.32.090194.001425},
  \href {http://adsabs.harvard.edu/abs/1994ARA%26A..32..277F} {32, 277}

\bibitem[\protect\citeauthoryear{{Fabian}}{{Fabian}}{2012}]{2012ARA&A..50..455F}
{Fabian} A.~C.,  2012, \mn@doi [\araa] {10.1146/annurev-astro-081811-125521},
  \href {http://adsabs.harvard.edu/abs/2012ARA%26A..50..455F} {50, 455}

\bibitem[\protect\citeauthoryear{{Fanaroff} \& {Riley}}{{Fanaroff} \&
  {Riley}}{1974}]{1974MNRAS.167P..31F}
{Fanaroff} B.~L.,  {Riley} J.~M.,  1974, \mn@doi [\mnras]
  {10.1093/mnras/167.1.31P}, \href
  {http://adsabs.harvard.edu/abs/1974MNRAS.167P..31F} {167, 31P}

\bibitem[\protect\citeauthoryear{{Feretti} \& {Venturi}}{{Feretti} \&
  {Venturi}}{2002}]{2002ASSL..272..163F}
{Feretti} L.,  {Venturi} T.,  2002, in {Feretti} L.,  {Gioia} I.~M.,
  {Giovannini} G.,  eds,  Astrophysics and Space Science Library Vol. 272,
  Merging Processes in Galaxy Clusters. pp 163--195,
  \mn@doi{10.1007/0-306-48096-4_6}

\bibitem[\protect\citeauthoryear{{Feretti}, {Giovannini}, {Govoni}  \&
  {Murgia}}{{Feretti} et~al.}{2012}]{2012A&ARv..20...54F}
{Feretti} L.,  {Giovannini} G.,  {Govoni} F.,   {Murgia} M.,  2012, \mn@doi
  [\aapr] {10.1007/s00159-012-0054-z}, \href
  {http://adsabs.harvard.edu/abs/2012A%26ARv..20...54F} {20, 54}

\bibitem[\protect\citeauthoryear{{Ferrari}, {Arnaud}, {Ettori}, {Maurogordato}
  \& {Rho}}{{Ferrari} et~al.}{2006}]{2006A&A...446..417F}
{Ferrari} C.,  {Arnaud} M.,  {Ettori} S.,  {Maurogordato} S.,   {Rho} J.,
  2006, \mn@doi [\aap] {10.1051/0004-6361:20053946}, \href
  {http://adsabs.harvard.edu/abs/2006A%26A...446..417F} {446, 417}

\bibitem[\protect\citeauthoryear{{Govoni} et~al.,}{{Govoni}
  et~al.}{2019}]{2019Sci...364..981G}
{Govoni} F.,  et~al., 2019, \mn@doi [Science] {10.1126/science.aat7500}, \href
  {https://ui.adsabs.harvard.edu/abs/2019Sci...364..981G} {364, 981}

\bibitem[\protect\citeauthoryear{{Harwood} et~al.,}{{Harwood}
  et~al.}{2016}]{2016MNRAS.458.4443H}
{Harwood} J.~J.,  et~al., 2016, \mn@doi [\mnras] {10.1093/mnras/stw638}, \href
  {http://adsabs.harvard.edu/abs/2016MNRAS.458.4443H} {458, 4443}

\bibitem[\protect\citeauthoryear{{Harwood} et~al.,}{{Harwood}
  et~al.}{2017}]{2017MNRAS.469..639H}
{Harwood} J.~J.,  et~al., 2017, \mn@doi [\mnras] {10.1093/mnras/stx820}, \href
  {http://adsabs.harvard.edu/abs/2017MNRAS.469..639H} {469, 639}

\bibitem[\protect\citeauthoryear{{Henriksen}}{{Henriksen}}{1996}]{1996rftu.proc..587H}
{Henriksen} M.~J.,  1996, in {Zimmermann} H.~U.,  {Tr{\"u}mper} J.,   {Yorke}
  H.,  eds, Roentgenstrahlung from the Universe. pp 587--588

\bibitem[\protect\citeauthoryear{{Intema}}{{Intema}}{2014}]{2014arXiv1402.4889I}
{Intema} H.~T.,  2014, ArXiv e-prints:1402.4889, \href
  {http://adsabs.harvard.edu/abs/2014arXiv1402.4889I} {}

\bibitem[\protect\citeauthoryear{{Intema}, {van der Tol}, {Cotton}, {Cohen},
  {van Bemmel}  \& {R{\"o}ttgering}}{{Intema}
  et~al.}{2009}]{2009A&A...501.1185I}
{Intema} H.~T.,  {van der Tol} S.,  {Cotton} W.~D.,  {Cohen} A.~S.,  {van
  Bemmel} I.~M.,   {R{\"o}ttgering} H.~J.~A.,  2009, \mn@doi [\aap]
  {10.1051/0004-6361/200811094}, \href
  {http://adsabs.harvard.edu/abs/2009A%26A...501.1185I} {501, 1185}

\bibitem[\protect\citeauthoryear{{Intema}, {Jagannathan}, {Mooley}  \&
  {Frail}}{{Intema} et~al.}{2017}]{2017A&A...598A..78I}
{Intema} H.~T.,  {Jagannathan} P.,  {Mooley} K.~P.,   {Frail} D.~A.,  2017,
  \mn@doi [\aap] {10.1051/0004-6361/201628536}, \href
  {http://adsabs.harvard.edu/abs/2017A%26A...598A..78I} {598, A78}

\bibitem[\protect\citeauthoryear{{Lagan{\'a}}, {Lima Neto}, {Andrade-Santos}
  \& {Cypriano}}{{Lagan{\'a}} et~al.}{2008}]{Lagana08}
{Lagan{\'a}} T.~F.,  {Lima Neto} G.~B.,  {Andrade-Santos} F.,   {Cypriano}
  E.~S.,  2008, \mn@doi [\aap] {10.1051/0004-6361:20079168}, \href
  {http://adsabs.harvard.edu/abs/2008A%26A...485..633L} {485, 633}

\bibitem[\protect\citeauthoryear{{Lagan{\'a}}, {Durret}  \&
  {Lopes}}{{Lagan{\'a}} et~al.}{2019}]{Lagana19}
{Lagan{\'a}} T.~F.,  {Durret} F.,   {Lopes} P.~A.~A.,  2019, \mn@doi [\mnras]
  {10.1093/mnras/stz148}, \href
  {http://adsabs.harvard.edu/abs/2019MNRAS.484.2807L} {484, 2807}

\bibitem[\protect\citeauthoryear{{Mann} \& {Ebeling}}{{Mann} \&
  {Ebeling}}{2012}]{2012MNRAS.420.2120M}
{Mann} A.~W.,  {Ebeling} H.,  2012, \mn@doi [\mnras]
  {10.1111/j.1365-2966.2011.20170.x}, \href
  {http://adsabs.harvard.edu/abs/2012MNRAS.420.2120M} {420, 2120}

\bibitem[\protect\citeauthoryear{{Markevitch} \& {Vikhlinin}}{{Markevitch} \&
  {Vikhlinin}}{2001}]{2001ApJ...563...95M}
{Markevitch} M.,  {Vikhlinin} A.,  2001, \mn@doi [\apj] {10.1086/323831}, \href
  {http://adsabs.harvard.edu/abs/2001ApJ...563...95M} {563, 95}

\bibitem[\protect\citeauthoryear{{Markevitch} \& {Vikhlinin}}{{Markevitch} \&
  {Vikhlinin}}{2007}]{2007PhR...443....1M}
{Markevitch} M.,  {Vikhlinin} A.,  2007, \mn@doi [\physrep]
  {10.1016/j.physrep.2007.01.001}, \href
  {http://adsabs.harvard.edu/abs/2007PhR...443....1M} {443, 1}

\bibitem[\protect\citeauthoryear{{Maurogordato}, {Sauvageot}, {Bourdin},
  {Cappi}, {Benoist}, {Ferrari}, {Mars}  \& {Houairi}}{{Maurogordato}
  et~al.}{2011}]{2011A&A...525A..79M}
{Maurogordato} S.,  {Sauvageot} J.~L.,  {Bourdin} H.,  {Cappi} A.,  {Benoist}
  C.,  {Ferrari} C.,  {Mars} G.,   {Houairi} K.,  2011, \mn@doi [\aap]
  {10.1051/0004-6361/201014415}, \href
  {http://adsabs.harvard.edu/abs/2011A%26A...525A..79M} {525, A79}

\bibitem[\protect\citeauthoryear{{McNamara} \& {Nulsen}}{{McNamara} \&
  {Nulsen}}{2007}]{2007ARA&A..45..117M}
{McNamara} B.~R.,  {Nulsen} P.~E.~J.,  2007, \mn@doi [\araa]
  {10.1146/annurev.astro.45.051806.110625}, \href
  {http://adsabs.harvard.edu/abs/2007ARA%26A..45..117M} {45, 117}

\bibitem[\protect\citeauthoryear{{McNamara} \& {Nulsen}}{{McNamara} \&
  {Nulsen}}{2012}]{2012NJPh...14e5023M}
{McNamara} B.~R.,  {Nulsen} P.~E.~J.,  2012, \mn@doi [New Journal of Physics]
  {10.1088/1367-2630/14/5/055023}, \href
  {http://adsabs.harvard.edu/abs/2012NJPh...14e5023M} {14, 055023}

\bibitem[\protect\citeauthoryear{{Miniati}, {Ryu}, {Kang}, {Jones}, {Cen}  \&
  {Ostriker}}{{Miniati} et~al.}{2000}]{2000ApJ...542..608M}
{Miniati} F.,  {Ryu} D.,  {Kang} H.,  {Jones} T.~W.,  {Cen} R.,   {Ostriker}
  J.~P.,  2000, \mn@doi [\apj] {10.1086/317027}, \href
  {http://adsabs.harvard.edu/abs/2000ApJ...542..608M} {542, 608}

\bibitem[\protect\citeauthoryear{{Molnar}, {Hearn}, {Haiman}, {Bryan}, {Evrard}
   \& {Lake}}{{Molnar} et~al.}{2009}]{2009ApJ...696.1640M}
{Molnar} S.~M.,  {Hearn} N.,  {Haiman} Z.,  {Bryan} G.,  {Evrard} A.~E.,
  {Lake} G.,  2009, \mn@doi [\apj] {10.1088/0004-637X/696/2/1640}, \href
  {http://adsabs.harvard.edu/abs/2009ApJ...696.1640M} {696, 1640}

\bibitem[\protect\citeauthoryear{{Murgia}, {Govoni}, {Feretti}  \&
  {Giovannini}}{{Murgia} et~al.}{2010}]{2010A&A...509A..86M}
{Murgia} M.,  {Govoni} F.,  {Feretti} L.,   {Giovannini} G.,  2010, \mn@doi
  [\aap] {10.1051/0004-6361/200913414}, \href
  {http://adsabs.harvard.edu/abs/2010A%26A...509A..86M} {509, A86}

\bibitem[\protect\citeauthoryear{{Nurgaliev}, {McDonald}, {Benson}, {Miller},
  {Stubbs}  \& {Vikhlinin}}{{Nurgaliev} et~al.}{2013}]{2013ApJ...779..112N}
{Nurgaliev} D.,  {McDonald} M.,  {Benson} B.~A.,  {Miller} E.~D.,  {Stubbs}
  C.~W.,   {Vikhlinin} A.,  2013, \mn@doi [\apj] {10.1088/0004-637X/779/2/112},
  \href {http://adsabs.harvard.edu/abs/2013ApJ...779..112N} {779, 112}

\bibitem[\protect\citeauthoryear{{Parekh}, {van der Heyden}, {Ferrari}, {Angus}
   \& {Holwerda}}{{Parekh} et~al.}{2015}]{2015A&A...575A.127P}
{Parekh} V.,  {van der Heyden} K.,  {Ferrari} C.,  {Angus} G.,   {Holwerda} B.,
   2015, \mn@doi [\aap] {10.1051/0004-6361/201424123}, \href
  {http://adsabs.harvard.edu/abs/2015A%26A...575A.127P} {575, A127}

\bibitem[\protect\citeauthoryear{{Parekh}, {Durret}, {Padmanabh}  \&
  {Pandge}}{{Parekh} et~al.}{2017}]{2017MNRAS.470.3742P}
{Parekh} V.,  {Durret} F.,  {Padmanabh} P.,   {Pandge} M.~B.,  2017, \mn@doi
  [\mnras] {10.1093/mnras/stx1457}, \href
  {http://cdsads.u-strasbg.fr/abs/2017MNRAS.470.3742P} {470, 3742}

\bibitem[\protect\citeauthoryear{{Paterno-Mahler}, {Blanton}, {Brodwin},
  {Ashby}, {Golden-Marx}, {Decker}, {Wing}  \& {Anand}}{{Paterno-Mahler}
  et~al.}{2017}]{2017ApJ...844...78P}
{Paterno-Mahler} R.,  {Blanton} E.~L.,  {Brodwin} M.,  {Ashby} M.~L.~N.,
  {Golden-Marx} E.,  {Decker} B.,  {Wing} J.~D.,   {Anand} G.,  2017, \mn@doi
  [\apj] {10.3847/1538-4357/aa7b89}, \href
  {http://adsabs.harvard.edu/abs/2017ApJ...844...78P} {844, 78}

\bibitem[\protect\citeauthoryear{{Peterson} \& {Fabian}}{{Peterson} \&
  {Fabian}}{2006}]{2006PhR...427....1P}
{Peterson} J.~R.,  {Fabian} A.~C.,  2006, \mn@doi [\physrep]
  {10.1016/j.physrep.2005.12.007}, \href
  {http://adsabs.harvard.edu/abs/2006PhR...427....1P} {427, 1}

\bibitem[\protect\citeauthoryear{{Pfrommer}, {Springel}, {En{\ss}lin}  \&
  {Jubelgas}}{{Pfrommer} et~al.}{2006}]{2006MNRAS.367..113P}
{Pfrommer} C.,  {Springel} V.,  {En{\ss}lin} T.~A.,   {Jubelgas} M.,  2006,
  \mn@doi [\mnras] {10.1111/j.1365-2966.2005.09953.x}, \href
  {http://adsabs.harvard.edu/abs/2006MNRAS.367..113P} {367, 113}

\bibitem[\protect\citeauthoryear{{Piffaretti}, {Arnaud}, {Pratt},
  {Pointecouteau}  \& {Melin}}{{Piffaretti} et~al.}{2011}]{2011A&A...534A.109P}
{Piffaretti} R.,  {Arnaud} M.,  {Pratt} G.~W.,  {Pointecouteau} E.,   {Melin}
  J.-B.,  2011, \mn@doi [\aap] {10.1051/0004-6361/201015377}, \href
  {http://adsabs.harvard.edu/abs/2011A%26A...534A.109P} {534, A109}

\bibitem[\protect\citeauthoryear{{Planck Collaboration} et~al.,}{{Planck
  Collaboration} et~al.}{2013}]{2013A&A...550A.134P}
{Planck Collaboration} et~al., 2013, \mn@doi [\aap]
  {10.1051/0004-6361/201220194}, \href
  {http://adsabs.harvard.edu/abs/2013A%26A...550A.134P} {550, A134}

\bibitem[\protect\citeauthoryear{{Planck Collaboration} et~al.,}{{Planck
  Collaboration} et~al.}{2014}]{2014A&A...571A..29P}
{Planck Collaboration} et~al., 2014, \mn@doi [\aap]
  {10.1051/0004-6361/201321523}, \href
  {https://ui.adsabs.harvard.edu/abs/2014A%26A...571A..29P} {571, A29}

\bibitem[\protect\citeauthoryear{{Poole}, {Fardal}, {Babul}, {McCarthy},
  {Quinn}  \& {Wadsley}}{{Poole} et~al.}{2006}]{2006MNRAS.373..881P}
{Poole} G.~B.,  {Fardal} M.~A.,  {Babul} A.,  {McCarthy} I.~G.,  {Quinn} T.,
  {Wadsley} J.,  2006, \mn@doi [\mnras] {10.1111/j.1365-2966.2006.10916.x},
  \href {http://adsabs.harvard.edu/abs/2006MNRAS.373..881P} {373, 881}

\bibitem[\protect\citeauthoryear{{Pranger}, {B{\"o}hm}, {Ferrari},
  {Maurogordato}, {Benoist}, {H{\"o}ller}  \& {Schindler}}{{Pranger}
  et~al.}{2014}]{2014A&A...570A..40P}
{Pranger} F.,  {B{\"o}hm} A.,  {Ferrari} C.,  {Maurogordato} S.,  {Benoist} C.,
   {H{\"o}ller} H.,   {Schindler} S.,  2014, \mn@doi [\aap]
  {10.1051/0004-6361/201424727}, \href
  {http://cdsads.u-strasbg.fr/abs/2014A%26A...570A..40P} {570, A40}

\bibitem[\protect\citeauthoryear{{Pratt}, {Arnaud}, {Biviano}, {Eckert},
  {Ettori}, {Nagai}, {Okabe}  \& {Reiprich}}{{Pratt}
  et~al.}{2019}]{2019SSRv..215...25P}
{Pratt} G.~W.,  {Arnaud} M.,  {Biviano} A.,  {Eckert} D.,  {Ettori} S.,
  {Nagai} D.,  {Okabe} N.,   {Reiprich} T.~H.,  2019, \mn@doi [Space Science
  Reviews] {10.1007/s11214-019-0591-0}, \href
  {https://ui.adsabs.harvard.edu/abs/2019SSRv..215...25P} {215, 25}

\bibitem[\protect\citeauthoryear{{Read} \& {Ponman}}{{Read} \&
  {Ponman}}{2003}]{ReadPonman03}
{Read} A.~M.,  {Ponman} T.~J.,  2003, \mn@doi [\aap]
  {10.1051/0004-6361:20031099}, \href
  {http://adsabs.harvard.edu/abs/2003A%26A...409..395R} {409, 395}

\bibitem[\protect\citeauthoryear{{Rossetti} et~al.,}{{Rossetti}
  et~al.}{2016}]{2016MNRAS.457.4515R}
{Rossetti} M.,  et~al., 2016, \mn@doi [\mnras] {10.1093/mnras/stw265}, \href
  {https://ui.adsabs.harvard.edu/abs/2016MNRAS.457.4515R} {457, 4515}

\bibitem[\protect\citeauthoryear{{Ryu}, {Kang}, {Hallman}  \& {Jones}}{{Ryu}
  et~al.}{2003}]{2003ApJ...593..599R}
{Ryu} D.,  {Kang} H.,  {Hallman} E.,   {Jones} T.~W.,  2003, \mn@doi [\apj]
  {10.1086/376723}, \href {http://adsabs.harvard.edu/abs/2003ApJ...593..599R}
  {593, 599}

\bibitem[\protect\citeauthoryear{{Sarazin}}{{Sarazin}}{2002}]{2002ASSL..272....1S}
{Sarazin} C.~L.,  2002, in {Feretti} L.,  {Gioia} I.~M.,   {Giovannini} G.,
  eds,  Astrophysics and Space Science Library Vol. 272, Merging Processes in
  Galaxy Clusters. pp 1--38, \mn@doi{10.1007/0-306-48096-4_1}

\bibitem[\protect\citeauthoryear{{Scaife} \& {Heald}}{{Scaife} \&
  {Heald}}{2012}]{2012MNRAS.423L..30S}
{Scaife} A.~M.~M.,  {Heald} G.~H.,  2012, \mn@doi [\mnras]
  {10.1111/j.1745-3933.2012.01251.x}, \href
  {http://adsabs.harvard.edu/abs/2012MNRAS.423L..30S} {423, L30}

\bibitem[\protect\citeauthoryear{{Springel} et~al.,}{{Springel}
  et~al.}{2005}]{2005Natur.435..629S}
{Springel} V.,  et~al., 2005, \mn@doi [\nat] {10.1038/nature03597}, \href
  {http://adsabs.harvard.edu/abs/2005Natur.435..629S} {435, 629}

\bibitem[\protect\citeauthoryear{{Tempel}, {Stoica}, {Mart{\'{\i}}nez},
  {Liivam{\"a}gi}, {Castellan}  \& {Saar}}{{Tempel}
  et~al.}{2014}]{2014MNRAS.438.3465T}
{Tempel} E.,  {Stoica} R.~S.,  {Mart{\'{\i}}nez} V.~J.,  {Liivam{\"a}gi} L.~J.,
   {Castellan} G.,   {Saar} E.,  2014, \mn@doi [\mnras]
  {10.1093/mnras/stt2454}, \href
  {http://adsabs.harvard.edu/abs/2014MNRAS.438.3465T} {438, 3465}

\bibitem[\protect\citeauthoryear{{Ulmer} \& {Cruddace}}{{Ulmer} \&
  {Cruddace}}{1982}]{1982ApJ...258..434U}
{Ulmer} M.~P.,  {Cruddace} R.~G.,  1982, \mn@doi [\apj] {10.1086/160096}, \href
  {http://cdsads.u-strasbg.fr/abs/1982ApJ...258..434U} {258, 434}

\bibitem[\protect\citeauthoryear{{Vikhlinin} \& {Markevitch}}{{Vikhlinin} \&
  {Markevitch}}{2002}]{2002AstL...28..495V}
{Vikhlinin} A.~A.,  {Markevitch} M.~L.,  2002, \mn@doi [Astronomy Letters]
  {10.1134/1.1499173}, \href
  {http://adsabs.harvard.edu/abs/2002AstL...28..495V} {28, 495}

\bibitem[\protect\citeauthoryear{{Vikhlinin} et~al.,}{{Vikhlinin}
  et~al.}{2009}]{2009ApJ...692.1033V}
{Vikhlinin} A.,  et~al., 2009, \mn@doi [\apj] {10.1088/0004-637X/692/2/1033},
  \href {http://adsabs.harvard.edu/abs/2009ApJ...692.1033V} {692, 1033 (V09)}

\bibitem[\protect\citeauthoryear{{Wei{\ss}mann}, {B{\"o}hringer}, {{\v S}uhada}
   \& {Ameglio}}{{Wei{\ss}mann} et~al.}{2013}]{2013A&A...549A..19W}
{Wei{\ss}mann} A.,  {B{\"o}hringer} H.,  {{\v S}uhada} R.,   {Ameglio} S.,
  2013, \mn@doi [\aap] {10.1051/0004-6361/201219333}, \href
  {http://adsabs.harvard.edu/abs/2013A%26A...549A..19W} {549, A19}

\bibitem[\protect\citeauthoryear{{Werner}, {Finoguenov}, {Kaastra},
  {Simionescu}, {Dietrich}, {Vink}  \& {B{\"o}hringer}}{{Werner}
  et~al.}{2008}]{2008A&A...482L..29W}
{Werner} N.,  {Finoguenov} A.,  {Kaastra} J.~S.,  {Simionescu} A.,  {Dietrich}
  J.~P.,  {Vink} J.,   {B{\"o}hringer} H.,  2008, \mn@doi [\aap]
  {10.1051/0004-6361:200809599}, \href
  {http://adsabs.harvard.edu/abs/2008A%26A...482L..29W} {482, L29}

\bibitem[\protect\citeauthoryear{{Wilber} et~al.,}{{Wilber}
  et~al.}{2019}]{2019A&A...622A..25W}
{Wilber} A.,  et~al., 2019, \mn@doi [\aap] {10.1051/0004-6361/201833884}, \href
  {https://ui.adsabs.harvard.edu/abs/2019A&A...622A..25W} {622, A25}

\bibitem[\protect\citeauthoryear{{Wing} \& {Blanton}}{{Wing} \&
  {Blanton}}{2011}]{2011AJ....141...88W}
{Wing} J.~D.,  {Blanton} E.~L.,  2011, \mn@doi [\aj]
  {10.1088/0004-6256/141/3/88}, \href
  {http://adsabs.harvard.edu/abs/2011AJ....141...88W} {141, 88}

\bibitem[\protect\citeauthoryear{{van Weeren}, {de Gasperin}, {Akamatsu},
  {Br{\"u}ggen}, {Feretti}, {Kang}, {Stroe}  \& {Zandanel}}{{van Weeren}
  et~al.}{2019}]{2019SSRv..215...16V}
{van Weeren} R.~J.,  {de Gasperin} F.,  {Akamatsu} H.,  {Br{\"u}ggen} M.,
  {Feretti} L.,  {Kang} H.,  {Stroe} A.,   {Zandanel} F.,  2019, \mn@doi [\ssr]
  {10.1007/s11214-019-0584-z}, \href
  {http://adsabs.harvard.edu/abs/2019SSRv..215...16V} {215, 16}

\makeatother
\end{thebibliography}
\newpage
\newpage
\appendix
\section*{Appendix}
\begin{figure}[htb]
    \centering
    \begin{subfigure}[t]{0.45\textwidth}
        \includegraphics[width=1\textwidth]{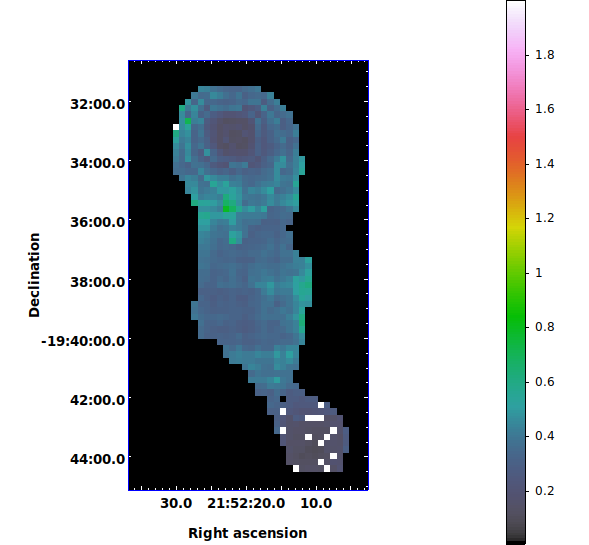}
        \caption{}
        \label{rfidtest_xaxis1}
    \end{subfigure}
    \begin{subfigure}[t]{0.45\textwidth}
        \includegraphics[width=1\textwidth]{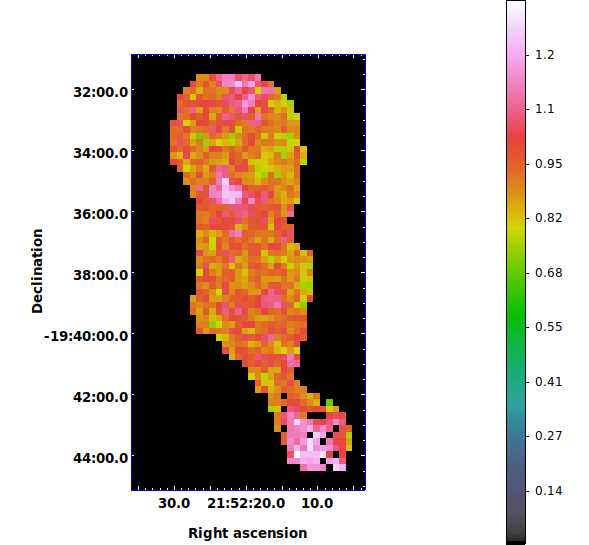}
        \caption{}
        \label{rfidtest_yaxis2}
        \end{subfigure}
    \caption{(a) XMM-{\it Newton} temperature error map in unit of keV. (b) Reduced $\chi^{2}$ map of temperature map.}
    \label{Xray_err_map}        
\end{figure}

\end{document}